\documentclass[aps,article,amsmath,amssymb,reprint,groupedaddress]{revtex4-2}

\usepackage{graphicx}
\usepackage{color}
\usepackage{dcolumn}
\usepackage{bm}
\usepackage[normalem]{ulem}
\usepackage{empheq}
\usepackage{here}
\usepackage{subcaption}
\usepackage{adjustbox}

\allowdisplaybreaks

\begin{document}
\title{Interatomic spin-orbit coupling in atomic orbital-based tight-binding models}
\author{Masaki Kato$^1$ and Masao Ogata$^{1,2}$}
\affiliation{$^1$Department of Physics, University of Tokyo, Hongo, Bunkyo-ku, Tokyo 113-0033, Japan \\
$^2$Trans-Scale Quantum Science Institute, University of Tokyo, Bunkyo, Tokyo 113-0033, Japan}

\date{\today}

\begin{abstract}
    Interatomic hopping mediated by spin-orbit coupling (SOC) entangles spin, orbital and sublattice degrees of freedom of electrons, leading to the emergence of intriguing phenomena such as novel topological insulators and exotic spin-dependent transport including chirality-induced spin selectivity (CISS). 
    Despite these effects, a comprehensive microscopic formalism to describe the spin-dependent hopping remains insufficiently established. 
    In this study, we systematically investigate SOC hopping by analytically deriving the hopping integrals within a two-center approximation based on atomic orbitals.
    Introducing independent parameters, or extended Slater-Koster symbols, that characterize SOC hopping, we explicitly determine the form of the hopping for $s$, $p$ and $d$ orbitals in the arbitrary hopping directions.
    Our formalism is then implemented in tight-binding models on several lattices.
    Furthermore, we examine the effect of SOC on band dispersion by employing a multipole decomposition for the SOC Hamiltonian, providing a fundamental understanding of SOC-induced phenomena.
    In particular, we derive an explicit expression for the SOC Hamiltonian that causes unique spin splitting in chiral systems by considering a triangular helical chain.
    Most importantly, the obtained SOC Hamiltonian does not contain a term that has the symmetry of electric toroidal monopole $G_0$ but rather an electric toroidal quadrupole $G_u$, which is the origin of chirality in this case.
\end{abstract}

\maketitle

\section{Introduction\label{sec:introduction}}
Spin-orbit coupling (SOC) is a relativistic correction to the Schr\"{o}dinger equation. 
Although this perturbative interaction is sometimes negligibly small, its involvement in the spin and orbital degrees of freedom of electrons can lead to exotic phenomena such as topological materials \cite{kane2005quantum, kane2005z, fu2007topological} and cross-correlated responses including magnetoelectric effect \cite{katsura2005spin, tokura2014multiferroics}.

The general expression for SOC is written as
\begin{align}
    H_{\mathrm{SOC}}=\frac{\hbar}{4m^2c^2}\bm{\sigma}\cdot(\bm{\nabla}V(\bm{r})\times\bm{p}),
    \label{eq:SOC}
\end{align}
where $m$ is the mass of an electron, $c$ is the speed of light, $\bm{p}$ is the momentum operator, $\bm{\sigma}$ is the Pauli matrix representing spin, and $V(\bm{r})$ is a periodic potential created by the nuclei.
$H_{\mathrm{SOC}}$ is difficult to deal with in a microscopic way because of $V(\bm{r})$.
If the potential is approximated as spherical, considering a single site, the SOC simplifies to the well-known form of $\bm{l}\cdot\bm{\sigma}$, where $\bm{l}$ represents the orbital angular momentum operator, resulting in a band splitting.

On the other hand, considering the contributions from two sites in $V(\bm{r})$, the SOC becomes an interaction between the orbital motion of electrons and their spin, manifesting itself in spin-dependent hopping integrals. 
Typical examples exhibiting this effect are spin Hall effect \cite{kato2004observation, PhysRevLett.94.047204, sinova2004universal} and the CISS effect \cite{ray1999asymmetric, gohler2011spin, kiran2016helicenes, onuki2014chiral, inui2020chirality, shiota2021chirality, ohe2024chirality}.
In the spin Hall effect, the trajectory of electrons is curved by the SOC even without an external magnetic field and the direction of motion depends on the spin orientation. 
Consequently, spin current flows perpendicularly to the applied elctric field.
The CISS effect is a phenomenon in which electrons transmitted through a chiral material become spin polarized depending on the handedneess of the material.
Since its discovery in organic molecules \cite{ray1999asymmetric, gohler2011spin, kiran2016helicenes}, similar spin polarization phenomena have been observed in inorganic crystals such as CrNb$_3$S$_6$, NbSi$_2$, TaSi$_2$ and $\alpha$-quartz \cite{onuki2014chiral, inui2020chirality, shishido2021detection, shiota2021chirality, ohe2024chirality}.
While it has been pointed out that SOC with a symmetry unique to chiral systems is crucial for a theoretical understanding of the CISS in metallic crystals \cite{suzuki2023spin}, its microscopic mechanism remains under debate.

For simple theoretical models used to describe spin-dependent hoppings, tight-binding models whose hopping integrals depend on the electron spin are often employed \cite{kane2005quantum, kane2005z, huertas2006spin, konschuh2010tight, yoda2015current, geyer2019chirality, hirakida2022chirality}. 
In most cases, however, the spin-dependent hopping is introduced phenomenologically in the Hamiltonian despite the fact that it should be derived from the microscopic Hamiltonian Eq. (\ref{eq:SOC}).
In the first half of this paper, we systematically investigate how the spin-dependent hopping can be derived microscopically.
Thereby, it is demonstrated that the spin-dependent hopping is determined only from the symmetry of the system and that of the orbitals involved in the hopping.
The method of deriving hopping integrals in a microscopic way has already been established for spin-independent hopping by Slater and Koster \cite{slater1954simplified} and subsequent studies \cite{sharma1979general, takegahara1980slater}.
In the present paper, we determine spin-dependent hopping within a two-center approximation by following Slater and Koster and discuss its parametrization, which we call extended Slater-Koster symbol for SOC.
We present the list of explicit expressions for $s, p$ and $d$ orbitals as examples.
Then, using the obtained spin-dependent hoppings and extended Slater-Koster symbols, we construct tight-binding models on several lattices, for example, square lattice, honeycomb lattice, and triangular helical chain.

Furthermore, we perform a multipole decomposition for the SOC Hamiltonians on these lattices and show that the SOC Hamiltonian can be classified as the product of three types of multipoles as discribed later.
Based on this classification, we argue SOC-induced band deformation and spin splitting.
Most interestingly, in the case of the triangular helical chain, we identify an SOC term that causes spin splitting typical in chiral systems.

There is a close relationship between SOC and the symmety of the system. 
It has long been recognized that particular types of SOC arise due to crystal symmetry such as the Rashba-type \cite{rashba1960properties} or the Dresselhaus-type \cite{dresselhaus1955spin}.
In the absence of spatial inversion symmetry, the SOC can result in an antisymmetric spin splitting.
With the presence of such momentum-spin hybridization, for example, superconductors can no longer be classified as singlet or triplet states, leading to the emergence of a novel superconducting state \cite{gor2001superconducting, frigeri2004superconductivity}.
On the other hand, the symmetry of the electronic degrees of freedom can be effectively described by multipoles. 
The concept of multipoles not only allows for the systematic classification of ordered states in electronic systems but also predicts cross-correlated responses emerging from those orders \cite{hayami2018classification, hayami2018microscopic, yatsushiro2021multipole}. 
Traditionally, the description of microscopic electronic states using the multipoles has been limited to orbitals with the same orbital angular momentum on a localized single atom. 
However, in recent years, extended multipoles have made it possible to go beyond the conventional picture and describe more diverse electronic states.
These extended multipoles include hybrid multipoles \cite{hayami2018microscopic, kusunose2020complete}, which are multipoles between states with different orbital angular momenta, and bond multipoles and cluster multipoles \cite{suzuki2018first, suzuki2017cluster, suzuki2019multipole, hayami2020bottom, yanagi2023generation}, which describe intersite or intrasite electronic degrees of freedom over several atoms.
 
This paper is organized as follows.
In Secs. \ref{sec:formulation} and \ref{sec:ESK_symbols}, we present the formulation for parametrizing interatomic SOC following Slater-Koster's method \cite{slater1954simplified} and provide a summary of the results obtained for the parametrization of SOC hopping between $s, p$, and $d$ orbitals within a two-center approximation.
In Sec. \ref{sec:TBmodels}, we show several examples of constructing tight-binding Hamiltonians taking into account the two-center interatomic SOC on specific lattices. 
Additionally, we analyze the symmetry of the SOC using the extended multipoles.
Section \ref{sec:summary} is devoted to the summary.

\section{Spin-dependent hopping in a two-center approximation\label{sec:formulation}}
We start from a general Hamiltonian with SOC,
\begin{align}
    H=\frac{\bm{p}^2}{2m}+V(\bm{r})+H_{\mathrm{SOC}}.
\end{align}
To construct a tight-binding model, we use the linear combination of atomic orbitals (LCAO) as a Bloch wave function: 
\begin{align}
    \sum_{k,l,\mu}a_{l\mu}\varphi_{l\mu}(\bm{r}-\bm{R}_k)e^{i\bm{k}\cdot\bm{R}_k},
\end{align}
where $\varphi_{l\mu}(\bm{r}-\bm{R}_k)$ is an atomic orbital localized at the $k$-th site $\bm{R}_k$ with quantum numbers $n, l, \mu$, and $a_{l\mu}$ is the corresponding coefficient.
In the present paper, we do not explicitly show the principal quantum number $n$. $l$ and $\mu= -l,-l+1,\cdots,l$ are the quantum numbers of tesseral harmonics $\mathcal{Y}_{l\mu}$, which are real functions created from a linear combination of spherical harmonics $Y_{lm}$. 
As an example, for $l=1$, $\mu=-1,0,1$ corresponds to $p_y,p_z$ and $p_x$ orbitals, respectively.
Later, we will explicitly relate $\mathcal{Y}_{l\mu}$ and $Y_{lm}$.

The hopping integral between two atomic orbitals $\varphi_{l\mu}(\bm{r}-\bm{R}_k)$ and $\varphi_{l'\mu'}(\bm{r}-\bm{R}_{k'})$ is given by
\begin{align}
    \int d\bm{r} \varphi_{l\mu}(\bm{r}-\bm{R}_k)H\varphi_{l'\mu'}(\bm{r}-\bm{R}_{k'}).
\end{align}
When we substitute $\frac{\bm{p}^2}{2m}+V(\bm{r})$ into $H$, we obtain the results by Slater and Koster \cite{slater1954simplified}. 
In the following, we focus on the spin-dependent hopping integral
\begin{align}
    \int d\bm{r} \varphi_{l\mu}(\bm{r}-\bm{R}_k)H_{\mathrm{SOC}}\varphi_{l'\mu'}(\bm{r}-\bm{R}_{k'}).
\end{align}
By substituting Eq. (\ref{eq:SOC}) and making an integration by parts, we can see that the spin-dependent hopping integral is given by
\begin{align}
    \label{eq:SOC_hopping_integral}
    i\frac{\hbar^2}{4m^2c^2}\bm{\lambda}_{l\mu;l'\mu'}(\bm{R}_k,\bm{R}_{k'})\cdot\bm{\sigma},
\end{align}
with
\begin{align}
    \bm{\lambda}_{l\mu;l'\mu'}(\bm{R}_k,\bm{R}_{k'}):=\int d\bm{r} V(\bm{r})&\bm{\nabla}\varphi_{l\mu}(\bm{r}-\bm{R}_k) \nonumber \\
    &\times\bm{\nabla}\varphi_{l'\mu'}(\bm{r}-\bm{R}_{k'}).
    \label{eq:lambda_def}
\end{align}

\begin{figure}[H]
    \begin{center}
        \includegraphics[width=8cm]{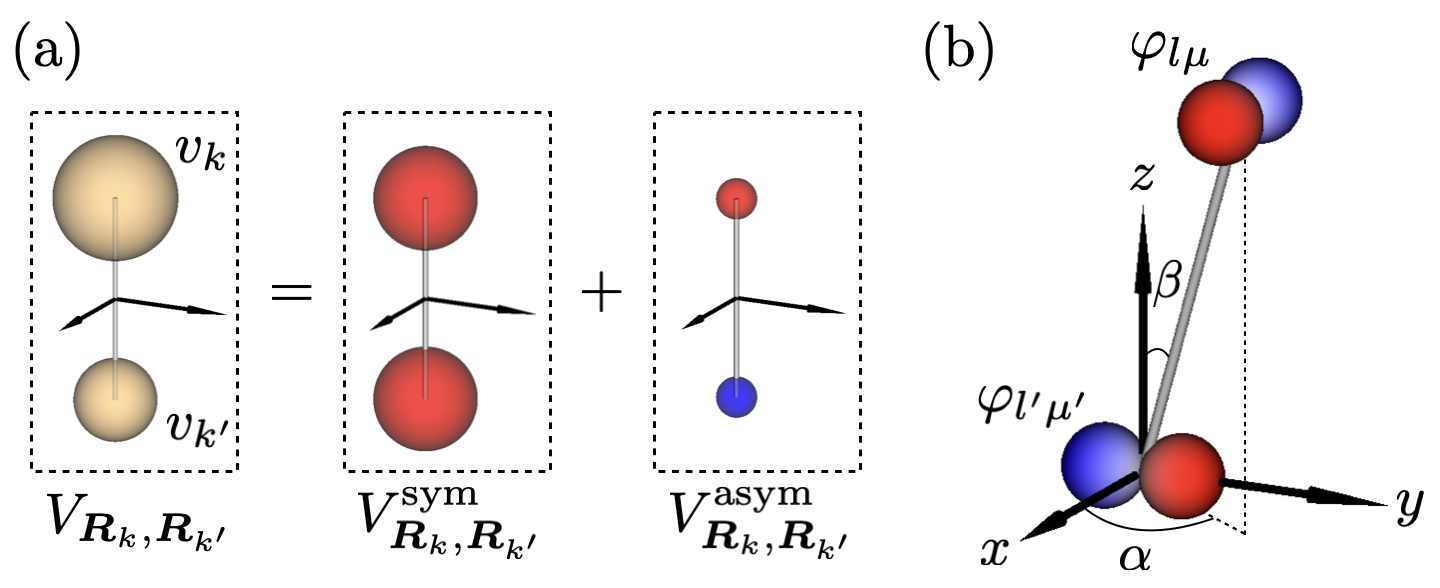}
        \caption{(a) In the two-center approximation, we only consider the sum of two atomic potentials $V_{\bm{R}_k,\bm{R}_{k'}}$, which can be decomposed into to symmetric and antisymmetric parts $V_{\bm{R}_k,\bm{R}_{k'}}^{\mathrm{sym}}$ and $V_{\bm{R}_k,\bm{R}_{k'}}^{\mathrm{asym}}$, respectively. The red and blue spheres represent positive and negative, respectively. (b) Visualization of hopping between atomic orbitals $\varphi_{l\mu}$ and $\varphi_{l'\mu'}$. The orientation of the hopping bond is specified using the polar angles $\alpha$ and $\beta$.}
        \label{fig:hopping}
    \end{center}
\end{figure}

In the following, we use a two-center approximation to simplify the integral following Slater and Koster \cite{slater1954simplified}, in which the potential $V(\bm{r})$ is reduced to $V_{\bm{R}_k,\bm{R}_{k'}}(\bm{r}) = v_k(\bm{r}-\bm{R}_k)+v_{k'}(\bm{r}-\bm{R}_{k'})$, where $v_k(\bm{r}-\bm{R}_k)$ and $v_{k'}(\bm{r}-\bm{R}_{k'})$ represent spherical atomic potentials for the two sites involved in the hopping. 
$v_k$ and $v_{k'}$ can be different in general if the two sites are inequivalent (see Fig. \ref{fig:hopping} (a)).
Thus, we decompose $V_{\bm{R}_k,\bm{R}_{k'}}$ into symmetric and antisymmetric parts, $V_{\bm{R}_k,\bm{R}_{k'}}^{\mathrm{sym}}$ and $V_{\bm{R}_k,\bm{R}_{k'}}^{\mathrm{asym}}$ defined by
\begin{align}
    V_{\bm{R}_k,\bm{R}_{k'}}^{\mathrm{sym}}(\bm{r}) = \frac{1}{2}{\left\{\right.}v_k(\bm{r}-\bm{R}_k)&+v_{k'}(\bm{r}-\bm{R}_{k'}) \nonumber \\
    + v_k(\bm{r}-\bm{R}_{k'})&+v_{k'}(\bm{r}-\bm{R}_k){\left.\right\}},
\end{align}
\begin{align}
    V_{\bm{R}_k,\bm{R}_{k'}}^{\mathrm{asym}}(\bm{r}) = \frac{1}{2}{\left\{\right.}v_k(\bm{r}-\bm{R}_k)&+v_{k'}(\bm{r}-\bm{R}_{k'}) \nonumber \\
    - v_k(\bm{r}-\bm{R}_{k'})&-v_{k'}(\bm{r}-\bm{R}_k){\left.\right\}},
\end{align}
which are symmetric and antisymmetric with respect to spatial inversion at the bond center, respectively.
Schematic pictures of $V_{\bm{R}_k,\bm{R}_{k'}}^{\mathrm{sym}}$ and $V_{\bm{R}_k,\bm{R}_{k'}}^{\mathrm{asym}}$ are shown in Fig. \ref{fig:hopping} (a).
By using $V_{\bm{R}_k,\bm{R}_{k'}}^{\mathrm{sym}}$ and $V_{\bm{R}_k,\bm{R}_{k'}}^{\mathrm{asym}}$, we define the symmetric and antisymmetric parts of Eq. (\ref{eq:lambda_def}) as
    \begin{align}
    \label{eq:lambda}
    \bm{\lambda}^{\mathrm{(a)sym}}_{l\mu;l'\mu'}(\bm{R}_k-\bm{R}_{k'}) = \int d\bm{r} V_{\bm{R}_k,\bm{R}_{k'}}^{\mathrm{(a)sym}}(\bm{r})&\bm{\nabla}\varphi_{l\mu}(\bm{r}-\bm{R}_k) \nonumber \\
    \times &\bm{\nabla}\varphi_{l'\mu'}(\bm{r}-\bm{R}_{k'}).
\end{align}
Since the hopping integral is independent of the position of the bond center, we have changed its augment to the relative position of the two sites $\bm{R}_k-\bm{R}_{k'}$.

The main purpose of the present paper is to classify $\bm{\lambda}^{\mathrm{(a)sym}}_{l\mu;l'\mu'}(\bm{R}_k-\bm{R}_{k'})$ based on the symmetries of the atomic orbitals $\varphi_{l\mu}$ and $\varphi_{l'\mu'}$. 
For this purpose, it is useful to use the spherical harmonics $Y_{lm}(\hat{\bm{r}})$ instead of the teesseral harmonics $\mathcal{Y}_{lm}(\hat{\bm{r}})$ with $l$ and $\mu$, where $\hat{\bm{r}}$ is a unit vector in the direction of $\bm{r}$.
The relationship between $\mathcal{Y}_{lm}(\hat{\bm{r}})$ and $Y_{lm}(\hat{\bm{r}})$ is
\begin{align}
    s \text{ wave}\hspace{1cm}& \mathcal{Y}_{00}(\hat{\bm{r}})=Y_{00}(\hat{\bm{r}}), \\
    \label{eq:p1}
    p \text{ wave}\hspace{1cm}& \mathcal{Y}_{11}(\hat{\bm{r}})=\frac{1}{\sqrt{2}}(-Y_{11}(\hat{\bm{r}})+Y_{1-1}(\hat{\bm{r}})), \\
    \label{eq:p2}
    & \mathcal{Y}_{10}(\hat{\bm{r}})=Y_{10}(\hat{\bm{r}}), \\
    \label{eq:p3}
    & \mathcal{Y}_{1-1}(\hat{\bm{r}})=\frac{i}{\sqrt{2}}(Y_{11}(\hat{\bm{r}})+Y_{1-1}(\hat{\bm{r}})), \\
    d \text{ wave}\hspace{1cm}& \mathcal{Y}_{22}(\hat{\bm{r}})=\frac{1}{\sqrt{2}}(Y_{22}(\hat{\bm{r}})+Y_{2-2}(\hat{\bm{r}})), \\
    & \mathcal{Y}_{21}(\hat{\bm{r}})=\frac{1}{\sqrt{2}}(-Y_{21}(\hat{\bm{r}})+Y_{2-1}(\hat{\bm{r}})), \\
    & \mathcal{Y}_{20}(\hat{\bm{r}})=Y_{20}(\hat{\bm{r}}), \\
    & \mathcal{Y}_{2-1}(\hat{\bm{r}})=\frac{i}{\sqrt{2}}(Y_{21}(\hat{\bm{r}})+Y_{2-1}(\hat{\bm{r}})), \\
    & \mathcal{Y}_{2-2}(\hat{\bm{r}})=\frac{i}{\sqrt{2}}(-Y_{22}(\hat{\bm{r}})+Y_{2-2}(\hat{\bm{r}})).
\end{align}
In general, we can write for $\mu=-l,-l+1,\cdots,l$
\begin{align}
    \label{eq:harmtotess}
    \mathcal{Y}_{l\mu}(\bm{\hat{r}})=\sum_{m=-l}^l C_{\mu m}Y_{lm}(\hat{\bm{r}}),
\end{align}
with
\begin{align}
    \label{eq:C1}
    C_{\mu\mu}=&\begin{cases}
        \frac{(-1)^{\mu}}{\sqrt{2}} & \text{for } \mu>0 \\
        1 & \text{for } \mu=0 \\
        \frac{i}{\sqrt{2}} & \text{for } \mu<0
    \end{cases}, \\
    \label{eq:C2}
    C_{\mu-\mu}=&\begin{cases}
        \frac{1}{\sqrt{2}} & \text{for } \mu>0 \\
        1 & \text{for } \mu=0 \\
        -\frac{(-1)^{\mu}i}{\sqrt{2}} & \text{for } \mu<0
    \end{cases}, \\
    \label{eq:C3}
    \text{Otherwise } &0.
\end{align}

Using the coefficients $C_{\mu m}$, the $\tau$-component ($\tau=x,y,z$) of Eq. (\ref{eq:lambda}) becomes 
\begin{align}
    \label{eq:lambda_CC}
    &\lambda^{\mathrm{(a)sym}}_{l\mu;l'\mu',\tau}(\bm{R}_k-\bm{R}_{k'}) = \sum_{\nu,\nu'}\sum_{m=-l}^{l}\sum_{m'=-l'}^{l'}\epsilon_{\nu\nu'\tau}C_{\mu m}C_{\mu' m'} \nonumber \\
    &\times\int d\bm{r} V_{\bm{R}_k,\bm{R}_{k'}}^{\mathrm{(a)sym}}(\bm{r})\partial_{\nu}\varphi_{lm}(\bm{r}-\bm{R}_k)\partial_{\nu'}\varphi_{l'm'}(\bm{r}-\bm{R}_{k'}),
\end{align}
where $\epsilon_{\nu\nu'\tau}$ denotes the Levi-Civita's totally antisymmetric symbol and the summation of $\nu$ and $\nu'$ are taken over $x,y$ and $z$.

Since the $\nu$ and $\nu'$ derivatives have the symmetry of the rank-1 tesseral harmonics, we can rewrite them in terms of the spherical harmonics using Eqs. (\ref{eq:p1})-(\ref{eq:p3}) or Eq. (\ref{eq:harmtotess}) with $l=1$.
In this respect, $x,y$ and $z$ appearing in the summation of $\nu$ and $\nu'$ can be regarded as $1,-1$ and $0$, respectively.
With this notation, Eq. (\ref{eq:lambda_CC}) can be written as
\begin{widetext}
    \begin{align}
        \label{eq:lambda_CCCC}
        \lambda^{\mathrm{(a)sym}}_{l\mu;l'\mu',\tau}(\bm{R}_k-\bm{R}_{k'}) = \sum_{\nu,\nu'}\sum_{m=-l}^{l}\sum_{m'=-l'}^{l'}\sum_{i=-1}^1&\sum_{i'=-1}^1\epsilon_{\nu\nu'\tau}C_{\mu m}C_{\mu' m'}C_{\nu i}C_{\nu'i'} \nonumber \\
        & \times\int d\bm{r} V_{\bm{R}_k,\bm{R}_{k'}}^{\mathrm{(a)sym}}(\bm{r})\partial_i\varphi_{lm}(\bm{r}-\bm{R}_k)\partial_{i'}\varphi_{l'm'}(\bm{r}-\bm{R}_{k'}).
    \end{align}
\end{widetext}

Finally, we perform rotational transformation to orient the hopping direction parallel to the quantum axis (the $z$-axis).
Let us determine the hopping direction $\bm{R}_k-\bm{R}_{k'}$ by polar angles $\alpha$ and $\beta$ as shown in Fig. \ref{fig:hopping} (b).
Then, Eq. (\ref{eq:lambda_CCCC}) becomes
\begin{widetext}
    \begin{align}
        \lambda_{l\mu;l'\mu',\tau}^{\mathrm{(a)sym}}(\bm{R}_k-\bm{R}_{k'}) = \sum_{\nu,\nu'}\sum_{m,n=-l}^{l}\sum_{m',n'=-l'}^{l'}\sum_{i,j=-1}^1\sum_{i',j'=-1}^1&\frac{1}{2}\epsilon_{\nu\nu'\tau}C_{\mu m}C_{\mu' m'}C_{\nu i}C_{\nu'i'} \nonumber \\
        &\times D_{ji}^{(1)}D_{j'i'}^{(1)}D_{nm}^{(l)}D_{n'm'}^{(l')}
        \begin{pmatrix}
            &l&l'& \\
            j&n&j'&n'
        \end{pmatrix}_{\mathrm{(a)sym}},
        \label{eq:lambda_formula}
    \end{align}
\end{widetext}
where $D^{(l)}=D^{(l)}(0,-\beta,-\alpha)$ is the Wigner D-matrix about the Euler angle $(-\alpha,-\beta,0)$ \cite{varshalovich1988quantum}.
The definition and the explicit expressions for the Wigner D-matrix are given in Appendix \ref{app:WDM}.
In Eq. (\ref{eq:lambda_formula}), we have defined extended Slater-Koster symbol for SOC as
\begin{widetext}
    \begin{align}
        \begin{pmatrix}
            &l&l'& \\
            j&n&j'&n'
        \end{pmatrix}_{\mathrm{(a)sym}}
        :=\int d\bm{r} V_{\frac{\bm{Z}}{2},-\frac{\bm{Z}}{2}}^{\mathrm{(a)sym}}(\bm{r})\left[\partial_j\varphi_{ln}\left(\bm{r}-\frac{\bm{Z}}{2}\right)\partial_{j'}\varphi_{l'n'}\left(\bm{r}+\frac{\bm{Z}}{2}\right) - (j\leftrightarrow j')\right],
        \label{eq:symbol}
    \end{align}
\end{widetext}
where $\bm{Z} = |\bm{R}_k-\bm{R}_{k'}|\bm{e}_z$ and $(j\leftrightarrow j')$ represents the first term with $j$ and $j'$ interchanged.
The factor $\frac{1}{2}$ in Eq. (\ref{eq:lambda_formula}) comes from the antisymmetrized definition of the symbol in Eq. (\ref{eq:symbol}).
We note that the newly defined symbol in Eq. (\ref{eq:symbol}) depends on the shape of the potential and the bond length, although they are not explicitly specified.

\section{Symmetries and explcit forms of extended Slater-Koster symbols for SOC\label{sec:ESK_symbols}}
In the following, we discuss the properties of the symbol defined by Eq. (\ref{eq:symbol}).
Since it is written in terms of the spherical harmonics, we can find several symmetries among the symbols.

First, $\phi$-rotation around the $z$-axis leads to $\varphi_{lm}\to e^{im\phi}\varphi_{lm}$.
Therefore, unless $j+n+j'+n'=0$,
\begin{align}
    \begin{pmatrix}
        &l&l'& \\
        j&n&j'&n'
    \end{pmatrix}_{\mathrm{(a)sym}}=0.
    \label{eq:property1}
\end{align}

Second, from the definition (\ref{eq:symbol}), 
\begin{align}
    \begin{pmatrix}
        &l&l'& \\
        j&n&j'&n'
    \end{pmatrix}_{\mathrm{(a)sym}}=-
    \begin{pmatrix}
        &l&l'& \\
        j'&n&j&n'
    \end{pmatrix}_{\mathrm{(a)sym}},
    \label{eq:property2}
\end{align}
which means that all symbols with $j=j'$ vanish, and we only have to consider $(j,j')=(\pm 1,0),(0,\pm 1),(\pm1,\mp1)$.
Combining with Eq. (\ref{eq:property1}), we can see that only $n'=-n\mp 1$ for $(j,j')=(\pm1,0), (0,\pm1)$ and $n'=-n$ for $(j,j')=(\pm 1,\mp 1)$ are allowed.

Third, taking the complex conjugate of Eq. (\ref{eq:symbol}), we obtain
\begin{align}
    \begin{pmatrix}
        &l&l'& \\
        j&n&j'&n'
    \end{pmatrix}_{\mathrm{(a)sym}}^{\scalebox{1}{$*$}}=
    \begin{pmatrix}
        &l&l'& \\
        \overline{j}&\overline{n}&\overline{j'}&\overline{n'}
    \end{pmatrix}_{\mathrm{(a)sym}},
    \label{eq:property3}
\end{align}
where $\overline{n}:=-n$ and $\overline{j}:=-j$.

Fourth, mirror reflection with respect to the $zx$ plane leads to $\varphi_{lm}\to(-1)^m\varphi_{l\overline{m}}=\varphi_{lm}^*$. Therefore we obtain
\begin{align}
    \begin{pmatrix}
        &l&l'& \\
        j&n&j'&n'
    \end{pmatrix}_{\mathrm{(a)sym}}=
    \begin{pmatrix}
        &l&l'& \\
        j&n&j'&n'
    \end{pmatrix}_{\mathrm{(a)sym}}^{\scalebox{1}{$*$}},
    \label{eq:property4}
\end{align}
which indicates that all the symbols are real.

Fifth, inversion leads to $\varphi_{lm}\to (-1)^l\varphi_{lm}$. We obtain
\begin{align}
    \begin{pmatrix}
        &l&l'& \\
        j&n&j'&n'
    \end{pmatrix}_{\mathrm{(a)sym}}&=(-1)^{l+l'+l''}
    \begin{pmatrix}
        &l'&l& \\
        j'&n'&j&n
    \end{pmatrix}_{\mathrm{(a)sym}}, 
    \label{eq:property5}
\end{align}
where $(-1)^{l''}=1$ for the symmetric case and $(-1)^{l''}=-1$ for the antisymmetric case.

\subsection{Cases with $l=l'$}
Let us first consder the case where the two orbitals have the same angular momentum, $l=l'$ and the same principal quantum number.
In this case, independent symbols are 
\begin{align}
    \begin{pmatrix}
        &l&l& \\
        1&n-1&0&\overline{n}
    \end{pmatrix}_{\mathrm{(a)sym}}
    \text{and }
    \begin{pmatrix}
        &l&l& \\
        1&n&\overline{1}&\overline{n}
    \end{pmatrix}_{\mathrm{sym}},
\end{align}
for $n=1,2,\cdots,l$. 
The other symbols are obtained from the properties Eqs. (\ref{eq:property1}) - (\ref{eq:property5}) as follows.

\begin{widetext}
    \begin{align}
        &\begin{pmatrix}
            &l&l& \\
            1&n-1&0&\overline{n}
        \end{pmatrix}_{\mathrm{(a)sym}}=
        -\begin{pmatrix}
            &l&l& \\
            0&n-1&1&\overline{n}
        \end{pmatrix}_{\mathrm{(a)sym}}=
        \begin{pmatrix}
            &l&l& \\
            \overline{1}&\overline{n-1}&0&n
        \end{pmatrix}_{\mathrm{(a)sym}}=
        -\begin{pmatrix}
            &l&l& \\
            0&\overline{n-1}&\overline{1}&n
        \end{pmatrix}_{\mathrm{(a)sym}} \nonumber \\
        &\hspace{1cm}=(-1)^{l''}\begin{pmatrix}
            &l&l& \\
            0&\overline{n}&1&n-1
        \end{pmatrix}_{\mathrm{(a)sym}}=
        -(-1)^{l''}\begin{pmatrix}
            &l&l& \\
            1&\overline{n}&0&n-1
        \end{pmatrix}_{\mathrm{(a)sym}} \nonumber \\
        &\hspace{1cm}=(-1)^{l''}\begin{pmatrix}
            &l&l& \\
            0&n&\overline{1}&\overline{n-1}
        \end{pmatrix}_{\mathrm{(a)sym}}=
        -(-1)^{l''}\begin{pmatrix}
            &l&l& \\
            \overline{1}&n&0&\overline{n-1}
        \end{pmatrix}_{\mathrm{(a)sym}} \\
        &\begin{pmatrix}
            &l&l& \\
            1&n&\overline{1}&\overline{n}
        \end{pmatrix}_{\mathrm{sym}}=
        -\begin{pmatrix}
            &l&l& \\
            \overline{1}&n&1&\overline{n}
        \end{pmatrix}_{\mathrm{sym}}=
        \begin{pmatrix}
            &l&l& \\
            \overline{1}&\overline{n}&1&n
        \end{pmatrix}_{\mathrm{sym}}=
        -\begin{pmatrix}
            &l&l& \\
            1&\overline{n}&\overline{1}&n
        \end{pmatrix}_{\mathrm{sym}}.
    \end{align}
\end{widetext}
Note that 
$\begin{pmatrix}
    &l&l& \\
    1&n&\overline{1}&\overline{n}
\end{pmatrix}_{\mathrm{asym}}$
should vanish because of Eqs. (\ref{eq:property3})-(\ref{eq:property5}).

\subsection{Cases with $l\neq l'$}
For the cases with $l<l'$, independent symbols are
\begin{align}
    \begin{pmatrix}
        &l&l'& \\
        1&n&0&\overline{n+1}
    \end{pmatrix}_{\mathrm{(a)sym}}
\end{align}
for $n=-l,-l+1,\cdots, l$, and
\begin{align}
    \begin{pmatrix}
        &l&l'& \\
        1&n&\overline{1}&\overline{n}
    \end{pmatrix}_{\mathrm{(a)sym}}
\end{align}
for $n=1,2,\cdots, l$.
The other symbols are obtained from
\begin{widetext}
    \begin{align}
        &\begin{pmatrix}
            &l&l'& \\
            1&n&0&\overline{n+1}
        \end{pmatrix}_{\mathrm{(a)sym}}=
        -\begin{pmatrix}
            &l&l'& \\
            0&n&1&\overline{n+1}
        \end{pmatrix}_{\mathrm{(a)sym}}=
        \begin{pmatrix}
            &l&l'& \\
            \overline{1}&\overline{n}&0&n+1
        \end{pmatrix}_{\mathrm{(a)sym}}=
        -\begin{pmatrix}
            &l&l'& \\
            0&\overline{n}&\overline{1}&n+1
        \end{pmatrix}_{\mathrm{(a)sym}} \; (-l\leq n\leq l) \\
        &\begin{pmatrix}
            &l&l'& \\
            1&n&\overline{1}&\overline{n}
        \end{pmatrix}_{\mathrm{(a)sym}}=
        -\begin{pmatrix}
            &l&l'& \\
            \overline{1}&n&1&\overline{n}
        \end{pmatrix}_{\mathrm{(a)sym}}=
        \begin{pmatrix}
            &l&l'& \\
            \overline{1}&\overline{n}&1&n
        \end{pmatrix}_{\mathrm{(a)sym}}=
        -\begin{pmatrix}
            &l&l'& \\
            1&\overline{n}&\overline{1}&n
        \end{pmatrix}_{\mathrm{(a)sym}} \; (1\leq n\leq l).
    \end{align}
\end{widetext}
The symbols with $l>l'$ are obtained from Eq. (\ref{eq:property5}).

\subsection{Independent parameters for hopping between the $s, p$ and $d$ orbitals}
\subsubsection{$s$-$s$ orbitals}
For this case, we can easily see that there is no non-zero symbols.

\subsubsection{$p$-$p$ orbitals}
There are three independent symbols, which are renamed as parameters in a way similar to Slater-Koster:
\begin{align}
    K_{pp\sigma}&:=
    \begin{pmatrix}
        &1&1& \\
        1&0&0&\overline{1}
    \end{pmatrix}_{\mathrm{sym}}, 
    K'_{pp\pi}:=
    \begin{pmatrix}
        &1&1& \\
        1&1&\overline{1}&\overline{1}
    \end{pmatrix}_{\mathrm{sym}}, \nonumber \\
    \tilde{K}_{pp\sigma}&:=
    \begin{pmatrix}
        &1&1& \\
        1&0&0&\overline{1}
    \end{pmatrix}_{\mathrm{asym}}.
\end{align}

\subsubsection{$d$-$d$ orbitals}
There are 6 independent symbols:
\begin{align}
    K_{dd\sigma}&:=
    \begin{pmatrix}
        &2&2& \\
        1&0&0&\overline{1}
    \end{pmatrix}_{\mathrm{sym}}, 
    K_{dd\pi}:=
    \begin{pmatrix}
        &2&2& \\
        1&1&0&\overline{2}
    \end{pmatrix}_{\mathrm{sym}}, \nonumber \\
    K'_{dd\pi}&:=
    \begin{pmatrix}
        &2&2& \\
        1&1&\overline{1}&\overline{1}
    \end{pmatrix}_{\mathrm{sym}}, 
    K'_{dd\delta}:=
    \begin{pmatrix}
        &2&2& \\
        1&2&\overline{1}&\overline{2}
    \end{pmatrix}_{\mathrm{sym}}, \nonumber \\
    \tilde{K}_{dd\sigma}&:=
    \begin{pmatrix}
        &2&2& \\
        1&0&0&\overline{1}
    \end{pmatrix}_{\mathrm{asym}},
    \tilde{K}_{dd\pi}:=
    \begin{pmatrix}
        &2&2& \\
        1&1&0&\overline{2}
    \end{pmatrix}_{\mathrm{asym}}.
\end{align}

\subsubsection{$s$-$p$ orbitals}
There are two independent symbols:
\begin{align}
    K_{sp\sigma}:=
    \begin{pmatrix}
        &0&1& \\
        1&0&0&\overline{1}
    \end{pmatrix}_{\mathrm{sym}},
    \tilde{K}_{sp\sigma}:=
    \begin{pmatrix}
        &0&1& \\
        1&0&0&\overline{1}
    \end{pmatrix}_{\mathrm{asym}}.
\end{align}

\subsubsection{$s$-$d$ orbitals}
There are two independent symbols:
\begin{align}
    K_{sd\sigma}:=
    \begin{pmatrix}
        &0&2& \\
        1&0&0&\overline{1}
    \end{pmatrix}_{\mathrm{sym}},
    \tilde{K}_{sd\sigma}:=
    \begin{pmatrix}
        &0&2& \\
        1&0&0&\overline{1}
    \end{pmatrix}_{\mathrm{asym}}.
\end{align}

\subsubsection{$p$-$d$ orbitals}
There are 8 independent symbols:
\begin{align}
    K_{pd\sigma}&:=
    \begin{pmatrix}
        &1&2& \\
        1&0&0&\overline{1}
    \end{pmatrix}_{\mathrm{sym}}, 
    K_{pd\pi}:=
    \begin{pmatrix}
        &1&2& \\
        1&1&0&\overline{2}
    \end{pmatrix}_{\mathrm{sym}}, \nonumber \\
    K_{pd\overline{\pi}}&:=
    \begin{pmatrix}
        &1&2& \\
        1&\overline{1}&0&0
    \end{pmatrix}_{\mathrm{sym}}, 
    K'_{pd\pi}:=
    \begin{pmatrix}
        &1&2& \\
        1&1&\overline{1}&\overline{1}
    \end{pmatrix}_{\mathrm{sym}}, \nonumber \\
    \tilde{K}_{pd\sigma}&:=
    \begin{pmatrix}
        &1&2& \\
        1&0&0&\overline{1}
    \end{pmatrix}_{\mathrm{asym}}, 
    \tilde{K}_{pd\pi}:=
    \begin{pmatrix}
        &1&2& \\
        1&1&0&\overline{2}
    \end{pmatrix}_{\mathrm{asym}}, \nonumber \\
    \tilde{K}_{pd\overline{\pi}}&:=
    \begin{pmatrix}
        &1&2& \\
        1&\overline{1}&0&0
    \end{pmatrix}_{\mathrm{asym}}, 
    \tilde{K}'_{pd\pi}:=
    \begin{pmatrix}
        &1&2& \\
        1&1&\overline{1}&\overline{1}
    \end{pmatrix}_{\mathrm{asym}}.
\end{align}

\subsection{Explicit form of the spin-dependent hopping}
Finally, we obtain the SOC part of the hopping integral by using Eq. (\ref{eq:lambda_formula}).
The angular dependence of the hopping bond $\bm{R}$, which appears in $D_{nm}^{(l)}$ etc. in Eq. (\ref{eq:lambda_formula}), is expressed using direction cosines $\hat{R}_x, \hat{R}_y$ and $\hat{R}_z$, which are defined by
\begin{align}
    \hat{R}_x&=\sin\beta\cos\alpha, \nonumber \\
    \hat{R}_y&=\sin\beta\sin\alpha, \nonumber \\
    \hat{R}_z&=\cos\beta.
\end{align}

There is a relation for the exchange of the orbitals:
\begin{align}
    \label{eq:orbital_exchange}
    \bm{\lambda}_{l\mu;l'\mu'}^{\mathrm{(a)sym}}(\bm{R})=-(-1)^{l+l'+l''}\bm{\lambda}_{l'\mu';l\mu}^{\mathrm{(a)sym}}(\bm{R}),
\end{align}
which can be derived by using Eqs. (\ref{eq:lambda_formula}) and (\ref{eq:property5}).
Eq. (\ref{eq:orbital_exchange}) suggests that the hopping integral between the same orbital i.e. $l=l'$ and $\mu=\mu'$ is zero for the symmetric potential case with $l''=0$.

\subsubsection{$s$-$s$ orbitals}
There is no SOC hopping integral between the $s$ orbitals.

\subsubsection{$p$-$p$ orbitals}
For the $p$-$p$ orbitals, we obtain for the symmetric potential case
\begin{align}
    \lambda_{y;z,y}^{\mathrm{sym}}(\bm{R})&=-(K_{pp\sigma}+K'_{pp\pi})\hat{R}_x\hat{R}_y, \\
    \lambda_{y;z,z}^{\mathrm{sym}}(\bm{R})&=-(K_{pp\sigma}+K'_{pp\pi})\hat{R}_x\hat{R}_z, \\
    \lambda_{y;z,x}^{\mathrm{sym}}(\bm{R})&=K_{pp\sigma}-(K_{pp\sigma}+K'_{pp\pi})\hat{R}_x^2.
\end{align}
The hopping integrals between the same orbitals are zero for the symmetric potential case as discussed before, that is,
\begin{align}
    \bm{\lambda}_{x;x}^{\mathrm{sym}}(\bm{R}) = \bm{\lambda}_{y;y}^{\mathrm{sym}}(\bm{R}) = \bm{\lambda}_{z;z}^{\mathrm{sym}}(\bm{R}) =0.
\end{align}
The other orbital components can be obtained by changing the coordinates in a cyclic manner and using Eq. (\ref{eq:orbital_exchange}).
Therefore, all the results for the $p$-$p$ orbitals can be expressed by a single equation: 
\begin{align}
    \label{eq:lambda_p}
    \bm{\lambda}_{\mu;\mu'}^{\mathrm{sym}}(\bm{R})=\sum_{\nu}&\epsilon_{\mu\mu'\nu}\left\{K_{pp\sigma}\bm{e}_{\nu}\vphantom{\sum_{\nu'}}\right. \nonumber \\
    &\left.-\sum_{\nu'}(K_{pp\sigma}+K'_{pp\pi})\hat{R}_{\nu}\hat{R}_{\nu'}\bm{e}_{\nu'}\right\}.
\end{align}
For the antisymmetric potential case, we have
\begin{align}
    \lambda_{y;y,y}^{\mathrm{asym}}(\bm{R})&=0, \\
    \lambda_{y;y,z}^{\mathrm{asym}}(\bm{R})&=2\tilde{K}_{pp\sigma}\hat{R}_x\hat{R}_y, \\
    \lambda_{y;y,x}^{\mathrm{asym}}(\bm{R})&=-2\tilde{K}_{pp\sigma}\hat{R}_y\hat{R}_z, \\ \nonumber \\
    \lambda_{y;z,y}^{\mathrm{asym}}(\bm{R})&=-\tilde{K}_{pp\sigma}\hat{R}_x\hat{R}_y, \\
    \lambda_{y;z,z}^{\mathrm{asym}}(\bm{R})&=\tilde{K}_{pp\sigma}\hat{R}_x\hat{R}_z, \\
    \lambda_{y;z,x}^{\mathrm{asym}}(\bm{R})&=\tilde{K}_{pp\sigma}(\hat{R}_y^2-\hat{R}_z^2),
\end{align}
which can be expressed by a single equation:
\begin{align}
    \lambda_{\mu;\mu',\tau}^{\mathrm{asym}}(\bm{R})=\tilde{K}_{pp\sigma}\sum_{\nu}\hat{R}_{\nu}(\epsilon_{\mu\tau\nu}\hat{R}_{\mu'}+\epsilon_{\mu'\tau\nu}\hat{R}_{\mu}).
\end{align}

\subsubsection{$s$-$p$ orbitals}
For the $s$-$p$ orbitals, we obtain for the symmetric potential case
\begin{align}
    \lambda_{s;y,y}^{\mathrm{sym}}(\bm{R})&=0, \\
    \lambda_{s;y,z}^{\mathrm{sym}}(\bm{R})&=K_{sp\sigma}\hat{R}_x, \\
    \lambda_{s;y,x}^{\mathrm{sym}}(\bm{R})&=-K_{sp\sigma}\hat{R}_z,
\end{align}
which can be exressed by a single equation:
\begin{align}
    \lambda_{s;\mu,\tau}^{\mathrm{sym}}(\bm{R})=K_{sp\sigma}\sum_{\nu}\epsilon_{\mu\tau\nu}\hat{R}_{\nu}.
\end{align}

For the antisymmetric potential case, the hopping integral has the same form as the symmetric potential case with a different parameter name:
\begin{align}
    \lambda_{s;y,y}^{\mathrm{asym}}(\bm{R})&=0, \\
    \lambda_{s;y,z}^{\mathrm{asym}}(\bm{R})&=\tilde{K}_{sp\sigma}\hat{R}_x, \\
    \lambda_{s;y,x}^{\mathrm{asym}}(\bm{R})&=-\tilde{K}_{sp\sigma}\hat{R}_z,
\end{align}
which can be exressed by a single equation:
\begin{align}
    \lambda_{s;\mu,\tau}^{\mathrm{asym}}(\bm{R})=\tilde{K}_{sp\sigma}\sum_{\nu}\epsilon_{\mu\tau\nu}\hat{R}_{\nu}.
\end{align}
We note that the symmetric and antisymmetric parts differ in sign when the orbitals are reversed, that is,
\begin{align}
    \lambda_{\mu;s,\tau}^{\mathrm{(a)sym}}(\bm{R})=(-1)^{l''}\lambda_{s;\mu,\tau}^{\mathrm{(a)sym}}(\bm{R}),
\end{align}
which is a special case of Eq. (\ref{eq:orbital_exchange}).

\subsection{$d$ orbitals}
In the similar way, the cases for $d$-$d$ orbitals, $s$-$d$ orbitals, and $p$-$d$ orbitals are obtained.
These results are provided in Appendix \ref{app:d-list}.

\section{Tight-binding model on specific lattices and multipole decomposition\label{sec:TBmodels}}
In this section, we present several examples illustrating the incorporation of interatomic SOC into a tight-binding model on several specific lattices. 
Hereafter, we focus on the $p$ orbitals on the lattices consisting of a single species of element and nearest-neighbor hopping.
We omit $\frac{\hbar^2}{4m^2c^2}$ in the SOC hopping integral in Eq. (\ref{eq:SOC_hopping_integral}), or equivalently assume it is included in the SOC hopping parameters.

The tight-binding Hamiltonian with interatomic SOC is given by
\begin{align}
    \label{eq:tight-binding}
    H&=H_{\mathrm{kin}}+H_{\mathrm{SOC}} \nonumber \\
    &=\sum_{j,j',\mu,\mu',\sigma}t_{\mu;\mu'}(\bm{R}_j-\bm{R}_{j'})c_{j\mu\sigma}^{\dagger}c_{j'\mu'\sigma}\nonumber \\
    &\quad + \sum_{j,j',\mu,\mu',\sigma,\sigma'}c_{j\mu\sigma}^{\dagger}\left[i\bm{\lambda}_{\mu;\mu'}^{\mathrm{sym}}(\bm{R}_j-\bm{R}_{j'})\cdot\bm{\sigma}\right]_{\sigma\sigma'}c_{j'\mu'\sigma'},
\end{align}
where the summations for $j$ and $j'$ are carried out across all lattice sites, $\mu,\mu'=x,y,z$, and $\sigma,\sigma'=\uparrow,\downarrow$. 
$c_{j\mu\sigma}^{\dagger}$ and $c_{j\mu\sigma}$ are the creation and annihilation operators for the $p_{\mu}$ orbital with spin $\sigma$ on the $j$-th site.
The spin-independent hopping $t_{\mu;\mu'}(\bm{R})$ is given by
\begin{align}
    \label{eq:t_parameter}
    t_{\mu;\mu'}(\bm{R})=V_{pp\pi}\delta_{\mu\mu'}+(V_{pp\sigma}-V_{pp\pi})\hat{R}_{\mu}\hat{R}_{\mu'},
\end{align} 
where $V_{pp\sigma}$ and $V_{pp\pi}$ are the Slater-Koster parameters for the $p$ orbitals. 
$\bm{\lambda}_{\mu;\mu'}^{\mathrm{sym}}(\bm{R})$ corresponds to the lambda vector in Eq. (\ref{eq:lambda_p}) derived for the $p$-$p$ orbitals in the previous section.
Since we assume that all lattice sites are composed of the same atom, the atomic potentials for all the sites are equivalent and thus only the symmetric part of the spin-dependent hopping integral appears.

It is convenient for the symmetry argument to express $\epsilon_{\mu\mu'\nu}$ in Eq. (\ref{eq:lambda_p}) by angular momentum.
Because the matrix elements of the orbital angular momentum operator $\bm{\ell}$ between $p_{\mu}$ and $p_{\mu'}$ orbitals is
\begin{align}
    \label{eq:OAM}
    (\bm{\ell})_{\mu\mu'}=-i\hbar\sum_{\tau}\bm{e}_{\tau}\epsilon_{\mu\mu'\tau},
\end{align}
we obtain
\begin{align}
    \label{eq:ils}
    i\bm{\lambda}_{\mu;\mu'}^{\mathrm{sym}}(\bm{R})\cdot\bm{\sigma} = &-\frac{1}{\hbar}\sum_{\nu,\nu'}(\ell_{\nu})_{\mu\mu'}\sigma_{\nu'}\left\{K_{pp\sigma}\delta_{\nu\nu'}\vphantom{\hat{R}_{\nu}}\right. \nonumber \\
    &\left.- (K_{pp\sigma}+K'_{pp\pi})\hat{R}_{\nu}\hat{R}_{\nu'}\right\},
\end{align}
where $\delta_{\nu\nu'}$ is the Kronecker delta.
As we will show later, this can be expressed as a tensor product of bond, orbital, and spin multipoles.

\subsection{$p$ orbitals on a square lattice (point group $D_{4h}$)}
\begin{figure}[t]
    \begin{center}
        \includegraphics[width=8.5cm]{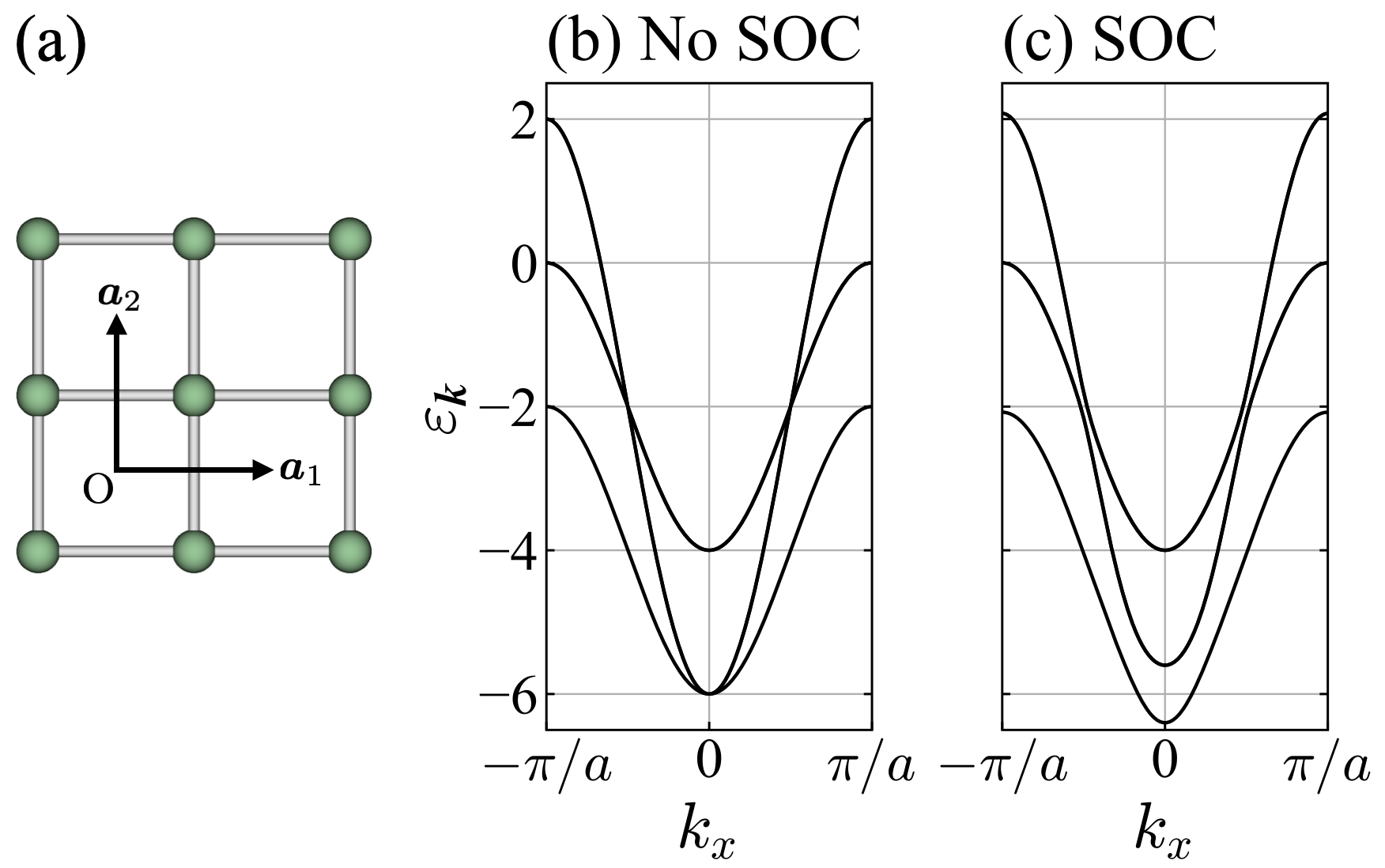}
        \caption{(a) Square lattice and its primitive translation vectors. (b)(c) Dispersion relations of the $p$ orbitals on a square lattice as a function of the wave vector $k_x$ with $k_y=k_z=0$ for the cases (b) without SOC and (c) with SOC. The Slater-Koster parameters for spin-independent hoppings are set as $V_{pp\sigma}=-2.0$ and $V_{pp\pi}=-1.0$. The SOC parameters for (c) are chosen as $K_{pp\sigma}=K_{pp\pi}'=0.1$.}
        \label{fig:disp_square}
    \end{center}
\end{figure}

First, we consider a square lattice shown in Fig. \ref{fig:disp_square} (a).
Let $\bm{a}_1=(a,0,0)^{\mathrm{T}}$ and $\bm{a}_2=(0,a,0)^{\mathrm{T}}$ be the primitive translation vectors.

Figure \ref{fig:disp_square} (b)(c) shows the band dispersion $\varepsilon_{\bm{k}}$ along $k_x$ with $k_y=0$ for the cases (b) without SOC and (c) with SOC. The Slater-Koster parameters are set as $V_{pp\sigma}=-2.0$ and $V_{pp\pi}=-1.0$. 
The SOC parameters for (c) are chosen as $K_{pp\sigma}=K_{pp\pi}'=0.1$.
As we can see from Fig. \ref{fig:disp_square} (c), the degeneracy at $k_x=0$ is lifted.
This symmetric band deformation around the $\Gamma$ point can be understood from a multipole decomposition of the Hamiltonian as follows \cite{hayami2018classification, hayami2020bottom}.

We investigate the symmetry of the obtained SOC Hamiltonian by performing a multipole decomposition of $H_{\mathrm{SOC}}$.
Generally, electronic Hamiltonians can be expanded with respect to four types of multipole operators, i.e., electric multipoles $Q_{lm}$, electric toroidal multipoles $G_{lm}$, magnetic multipoles $M_{lm}$ and magnetic toroidal multipoles $T_{lm}$, where $l$ and $m$ are the quantum numbers of the corresponding spherical harmonics $Y_{lm}$ \cite{hayami2018microscopic, watanabe2018group}.
In the following, we use conventional names for these multipoles by taking their linear combination.
The list of the conventional names up to rank-4 multipoles is given in Appendix \ref{app:convention}.

\begin{figure}[b]
    \begin{center}
        \includegraphics[width=7.5cm]{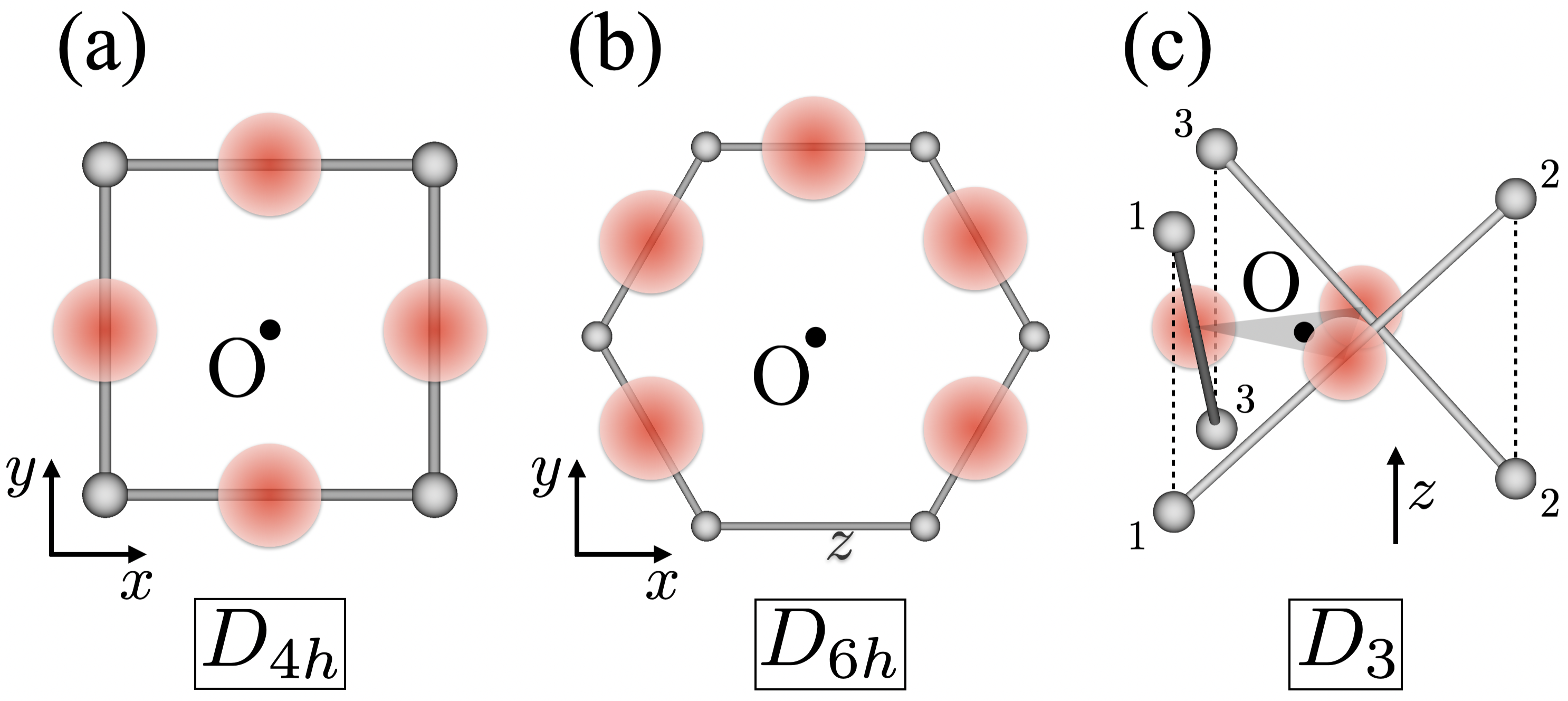}
        \caption{Clusters needed when discussing the symmetry of bond-dependent hopping integrals in the case of (a) a square lattice with point group $D_{4h}$, (b) a honeycomb lattice with point group $D_{6h}$ and  (c) a triangular helical chain with point group $D_3$.
        The red circles represent the position of ``charges'' at the bond centers.
        In (c), the three bonds consisting of a triangular helical chain ($1\to 2\to 3\to 1\to 2\to 3\to \cdots$) are shifted in the $z$-direction so that their bond centers are located in the same $xy$ plane, forming the gray shaded triangle.}
        \label{fig:clusters}
    \end{center}
\end{figure}

\begin{figure}[t]
    \begin{center}
        \includegraphics[width=7cm]{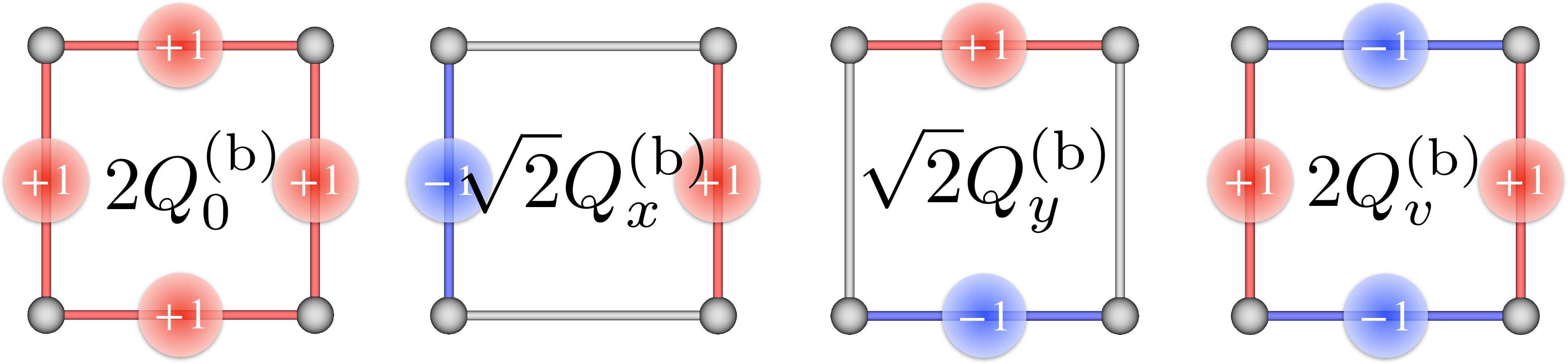}
        \caption{Bond multipoles for a square cluster. The red and blue bonds represent the presence of charges $+1$ and $-1$ at the bond centers, respectively. The gray bonds with no sphere represent the absence of a charge. For a square cluster, there are four bond multipoles and each charge configuration corresponds to an electric multipole represented by $Q^{(\mathrm{b})}_X$, where $X=0,x,y,v$ and the superscript (b) indicates it is a bond multipole.}
        \label{fig:square_bond}
    \end{center}
\end{figure}

The orbital angular momentum operator $\ell_{\nu}$ and the spin operator $\sigma_{\nu'}$  in Eq. (\ref{eq:ils}) are both represented by the magnetic dipoles $M_{\nu}^{(\mathrm{o})}$ and $M_{\nu'}^{(\mathrm{s})}$, respectively.
Here, the superscripts $(\mathrm{o})$ and $(\mathrm{s})$ indicate that the multipoles are derived from orbitals and spins, respectively.
On the other hand, any bond-dependent coefficient in the curly brackets in Eq. (\ref{eq:ils}) can be decomposed into the sum of irreducible representations of the point group of the lattice.
To analyze the symmetry of the bond-dependent hopping integral, we consider the cluster obtained by performing all the symmetry operations in the point group of the lattice to a given bond.
For the square lattice with point group $D_{4h}$, the cluster is shown in Fig. \ref{fig:clusters} (a), in which the origin of the symmetry operations is set to be the center of a plaquette.
Then, we assign ``charges'' on the four bond centers of this cluster depending on the value of the bond-dependent coefficient.
Any charge configuration on the square cluster can be expressed as a superposition of four basis configurations, which are chosen to have the symmetry of multipoles.
Figure \ref{fig:square_bond} illustrates the basis configurations for the square cluster and the corresponding multipoles are electric multipoles $Q_X$, where $X=0,x,y,v$ from left to right.
The red and blue spheres represent the presence of charges $+1$ and $-1$ at the bond centers, respectively.
These basis configurations are referred to as bond multipoles \cite{hayami2020bottom}.
To explicitly indicate they are bond multipoles, the superscript (b) is added to the multipoles.
We note that the coefficients of $Q_X$ written in Fig. \ref{fig:square_bond} are determined to normalize the bond multipoles such that the square root of the sum of squares of the charges on all bonds equals $1$.
In the case of $Q_x^{(\mathrm{b})}$, for example, $\sqrt{(+1)^2+0^2+(-1)^2+0^2}=\sqrt{2}$ is prefixed to $Q_x^{(\mathrm{b})}$.
The formal definition of the bond multipoles for a general $N$-bond cluster is given in Appendix \ref{app:bondmultipoles}.
If we consider complex hopping integrals, it is necessary to introduce $N$ additional multipole bases for imaginary hopping, which will not appear in the following discussion \cite{hayami2020bottom}.

Next, we analyze the multipoles of each term in $H_{\mathrm{SOC}}$ using these bond multipoles and magnetic dipoles for orbitals and spins.
The multipole corresponding to the first term in Eq. (\ref{eq:ils}), which does not depend on the bond angle, is 
\begin{align}
    \label{eq:square_multipoles1}
    \vcenter{\hbox{\includegraphics[width=1.2cm]{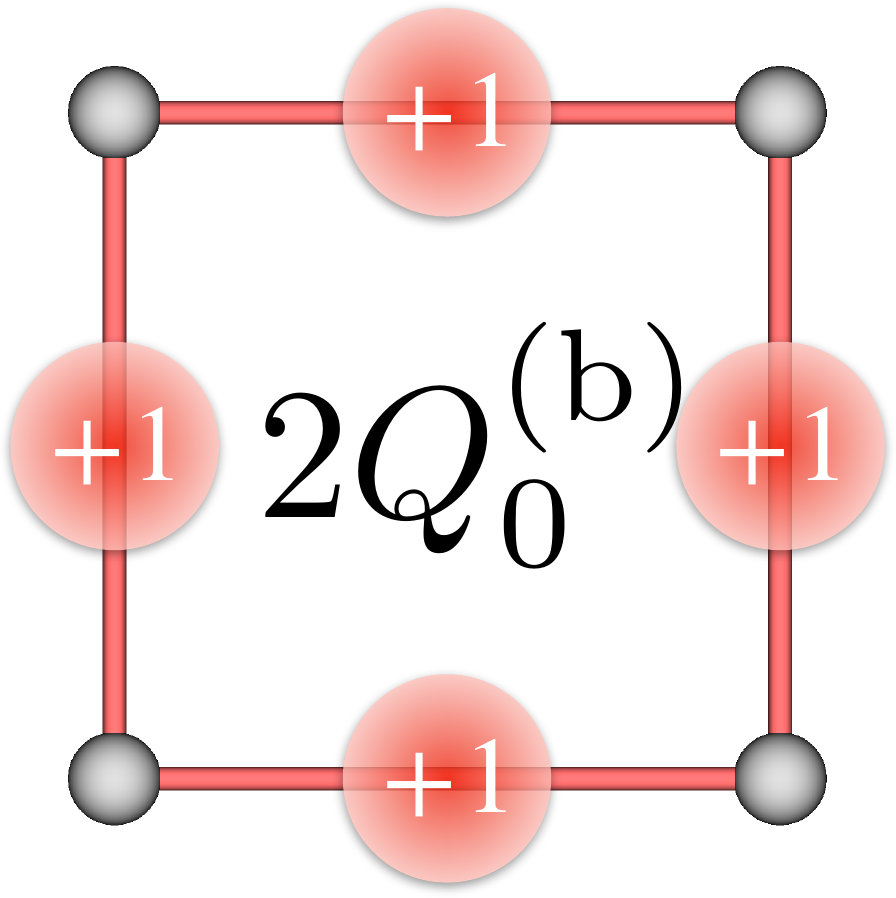}}}\otimes(\sum_{\nu=x,y,z}M_{\nu}^{(\mathrm{o})}\otimes M_{\nu}^{(\mathrm{s})}).
\end{align}

In the second term in Eq. (\ref{eq:ils}), $\hat{R}_{\nu}\hat{R}_{\nu'}$ depends on the bond.
For the bond in the $x$-direction, we have $\hat{\bm{R}}=\pm(1,0,0)^{\mathrm{T}}$ and in the $y$-direction, $\hat{\bm{R}}=\pm(0,1,0)^{\mathrm{T}}$.
Here, the double signs $\pm$ are determined by hopping in either the positive or negative direction.
Thus, for the $x$-directional bond, we have only $\nu=\nu'=x$, i.e., $M_x^{(\mathrm{o})}\otimes M_x^{(\mathrm{s})}$.
Similarly, for the $y$-directional bond, we have only $M_y^{(\mathrm{o})}\otimes M_y^{(\mathrm{s})}$.
From these, we can see that the multipole corresponding to this second term is
\begin{align}
    \label{eq:square_multipoles2}
    &\frac{1}{2}\vcenter{\hbox{\includegraphics[width=1.2cm]{square_Q0_v3.pdf}}}\otimes(M_x^{(\mathrm{o})}\otimes M_x^{(\mathrm{s})}+M_y^{(\mathrm{o})}\otimes M_y^{(\mathrm{s})}) \nonumber \\
    &-\frac{1}{2}\vcenter{\hbox{\includegraphics[width=1.2cm]{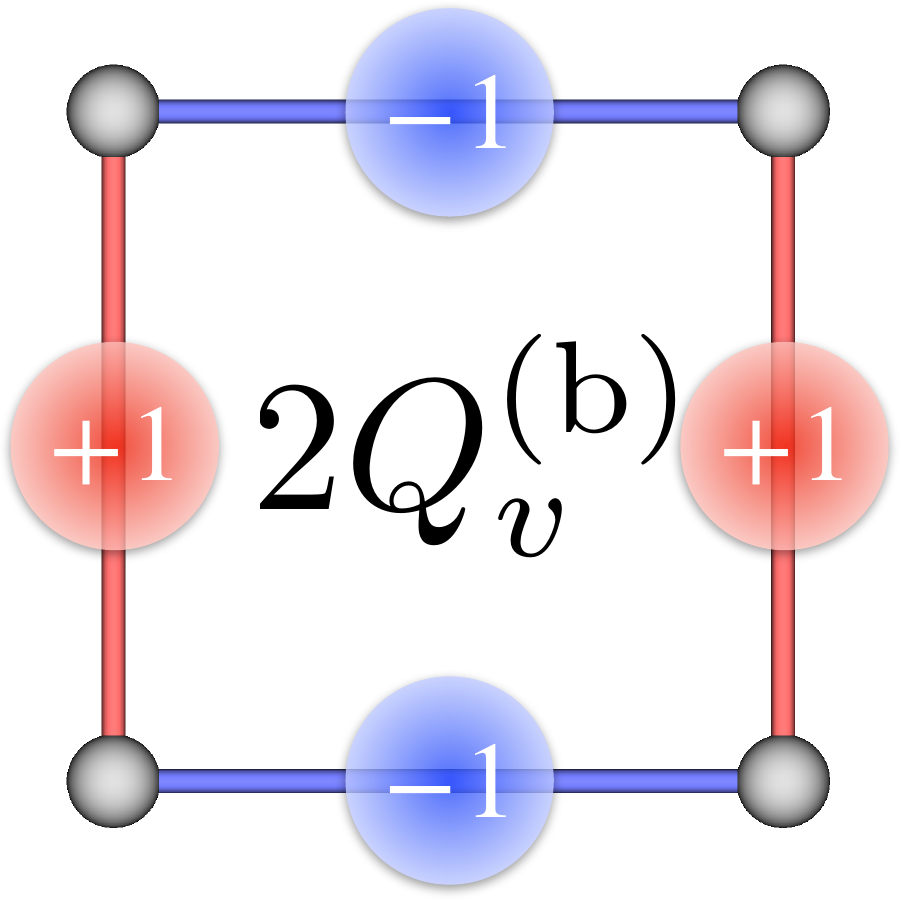}}}\otimes(M_x^{(\mathrm{o})}\otimes M_x^{(\mathrm{s})}-M_y^{(\mathrm{o})}\otimes M_y^{(\mathrm{s})}).
\end{align}

Then we expand Eqs. (\ref{eq:square_multipoles1}) and (\ref{eq:square_multipoles2}) in the basis of composite multipoles created by taking linear combination of the tensor product of the three multipoles $Q^{(\mathrm{b})}_X\otimes M^{(\mathrm{o})}_{\mu}\otimes M^{(\mathrm{s})}_{\nu}$ with appropriate coefficients in such a way that the new basis is symmetry-adapted \cite{kusunose2020complete, kusunose2023symmetry}.

Before discussing Eqs. (\ref{eq:square_multipoles1}) and (\ref{eq:square_multipoles2}), we explain the general procedure.
A new composite multipole $Z_{l_1l_2;lm}^{(\mathrm{AB})}$ can be constructed combining two types of multipoles by using
\begin{align}
    \label{eq:comp_multi}
    Z_{l_1l_2;lm}^{(\mathrm{AB})} = i^{l-l_1-l_2}&\sum_{m_1=-l_1}^{l_1}\sum_{m_2=-l_2}^{l_2} \nonumber \\
    &\langle l_1 m_1;l_2 m_2 | l m \rangle Z_{l_1m_1}^{(\mathrm{A})}\otimes Z_{l_2m_2}^{(\mathrm{B})},
\end{align}
where $\langle l_1 m_1;l_2 m_2 | l m \rangle$ is the Clebsch-Gordan coefficient.
$Z_{l_1m_1}^{(\mathrm{A})}$ and $Z_{l_2m_2}^{(\mathrm{B})}$ are multipoles having the symmetry of spherical harmonics $Y_{l_1m_1}$ and $Y_{l_2m_2}$, respectively, and $(\mathrm{A})$ and $(\mathrm{B})$ represent the types of the multipoles, which can be $(\mathrm{b})$, $(\mathrm{o})$ or $(\mathrm{s})$. 
The new multipole $Z_{l_1l_2;lm}^{(\mathrm{AB})}$ has the symmetry of $Y_{lm}$ and is labeled with $(\mathrm{AB})$, $l_1$ and $l_2$ to specify the original multipoles.
The prefactor $i^{l-l_1-l_2}$ is used to satisfy \cite{kusunose2020complete}
\begin{align}
    \left[ Z_{l_1l_2;lm}^{(\mathrm{AB})} \right]^{\dagger} = (-1)^m Z_{l_1l_2;l-m}^{(\mathrm{AB})}.
\end{align}
We may define the real form of $Z_{l_1l_2;lm}^{(\mathrm{AB})}$ similarly to Eq. (\ref{eq:harmtotess}) by using the coefficients $C_{\mu m}$ defined in Eqs. (\ref{eq:C1})-(\ref{eq:C3}).

Practically, we first create a composite multipole basis set by combining the multipoles for orbitals $M_{\nu}^{(\mathrm{o})}$ and those for spins $M_{\nu}^{(\mathrm{s})}$ as follows:
\begin{align}
    \label{eq:os_basis1}
    Q_0^{(\mathrm{os})}&=\frac{1}{\sqrt{3}}\sum_{\nu=x,y,z}M_{\nu}^{(\mathrm{o})}\otimes M_{\nu}^{(\mathrm{s})}, \\
    G_x^{(\mathrm{os})}&=\frac{1}{\sqrt{2}}(M_y^{(\mathrm{o})}\otimes M_z^{(\mathrm{s})} - M_z^{(\mathrm{o})}\otimes M_y^{(\mathrm{s})}), \\
    G_y^{(\mathrm{os})}&=\frac{1}{\sqrt{2}}(M_z^{(\mathrm{o})}\otimes M_x^{(\mathrm{s})} - M_x^{(\mathrm{o})}\otimes M_z^{(\mathrm{s})}), \\
    G_z^{(\mathrm{os})}&=\frac{1}{\sqrt{2}}(M_x^{(\mathrm{o})}\otimes M_y^{(\mathrm{s})} - M_y^{(\mathrm{o})}\otimes M_x^{(\mathrm{s})}), \\
    Q_u^{(\mathrm{os})}&=\frac{1}{\sqrt{6}}(3M_z^{(\mathrm{o})}\otimes M_z^{(\mathrm{s})}-\sum_{\nu=x,y,z}M_{\nu}^{(\mathrm{o})}\otimes M_{\nu}^{(\mathrm{s})}), \\
    Q_{zx}^{(\mathrm{os})}&=\frac{1}{\sqrt{2}}(M_z^{(\mathrm{o})}\otimes M_x^{(\mathrm{s})} + M_x^{(\mathrm{o})}\otimes M_z^{(\mathrm{s})}), \\
    Q_{yz}^{(\mathrm{os})}&=\frac{1}{\sqrt{2}}(M_y^{(\mathrm{o})}\otimes M_z^{(\mathrm{s})} + M_z^{(\mathrm{o})}\otimes M_y^{(\mathrm{s})}), \\
    Q_v^{(\mathrm{os})}&=\frac{1}{\sqrt{2}}(M_x^{(\mathrm{o})}\otimes M_x^{(\mathrm{s})} - M_y^{(\mathrm{o})}\otimes M_y^{(\mathrm{s})}), \\
    \label{eq:os_basis9}
    Q_{xy}^{(\mathrm{os})}&=\frac{1}{\sqrt{2}}(M_x^{(\mathrm{o})}\otimes M_y^{(\mathrm{s})} + M_y^{(\mathrm{o})}\otimes M_x^{(\mathrm{s})}).
\end{align}
Here, $Q_0^{(\mathrm{os})}$ is an electric monopole, $G_x^{(\mathrm{os})}, G_y^{(\mathrm{os})}, G_z^{(\mathrm{os})}$ are electric toroidal dipoles, and $Q_u^{(\mathrm{os})}, Q_{zx}^{(\mathrm{os})}, Q_{yz}^{(\mathrm{os})}, Q_v^{(\mathrm{os})}, Q_{xy}^{(\mathrm{os})}$ are electric quadrupoles, respectively.
Then, applying Eq. (\ref{eq:comp_multi}) to ${Q_X^{(\mathrm{b})}}$ and Eqs. (\ref{eq:os_basis1})-(\ref{eq:os_basis9}), we obtain all the multipole bases composed of three types of multipoles, which we label with $(\mathrm{bos})$.

In the case of Eq. (\ref{eq:square_multipoles1}), we have $Q_0^{(\mathrm{b})}\otimes Q_0^{(\mathrm{os})}=Q_0^{(\mathrm{bos})}$.
In the case of Eq. (\ref{eq:square_multipoles2}), the first term is proportional to 
\begin{align}
    Q_0^{(\mathrm{b})}\otimes \left(\sqrt{3}Q_0^{(\mathrm{os})}-\frac{\sqrt{6}}{2}Q_u^{(\mathrm{os})}\right),
\end{align}
which contains $Q_0^{(\mathrm{bos})}$ and $Q_u^{(\mathrm{bos})}$.
The second term is written as 
\begin{align}
    Q_v^{(\mathrm{b})}\otimes \sqrt{2}Q_v^{(\mathrm{os})},
\end{align}
which is decomposed as a sum of $Q_0^{(\mathrm{bos})}, Q_u^{(\mathrm{bos})}, Q_4^{(\mathrm{bos})}$ and $Q_{4u}^{(\mathrm{bos})}$.

Finally, the electronic band dispersion $\varepsilon_{n\bm{k}\sigma}$, where $n$ and $\sigma$ represent the band index and the spin polarization, is transformed by spatial inversion $\mathcal{P}$ and time reversal $\mathcal{T}$ as follows:
\begin{align}
    \label{eq:SIS}
    \varepsilon_{n\bm{k}\sigma} &\xrightarrow{\mathcal{P}} \varepsilon_{n'\bm{-k}\sigma} \\
    \label{eq:TRS}
    \varepsilon_{n\bm{k}\sigma} &\xrightarrow{\mathcal{T}} \varepsilon_{\overline{n}\bm{-k}-\sigma}.
\end{align}
Here, the band index $n$ is assumed to change to $n'$ and $\overline{n}$ by $\mathcal{P}$ and $\mathcal{T}$, respetively.
Since all the multipoles appearing in $H_{\mathrm{SOC}}$ are even-rank electric multipoles, the system preserves both the spatial inversion symmetry and the time reversal symmetry.
Therefore, $\varepsilon_{n\bm{k}\sigma}=\varepsilon_{\overline{n}'\bm{k}-\sigma}=\varepsilon_{n'\bm{-k}\sigma}=\varepsilon_{\overline{n}\bm{-k}-\sigma}$, indicating that the SOC results in the symmetric band deformation without spin-splitting or antisymmetric band deformation \cite{hayami2018classification, hayami2020bottom}.

\subsection{$p$ orbitals on a honeycomb lattice (point group $D_{6h}$)}
\begin{figure}[t]
    \begin{center}
        \includegraphics[width=8.5cm]{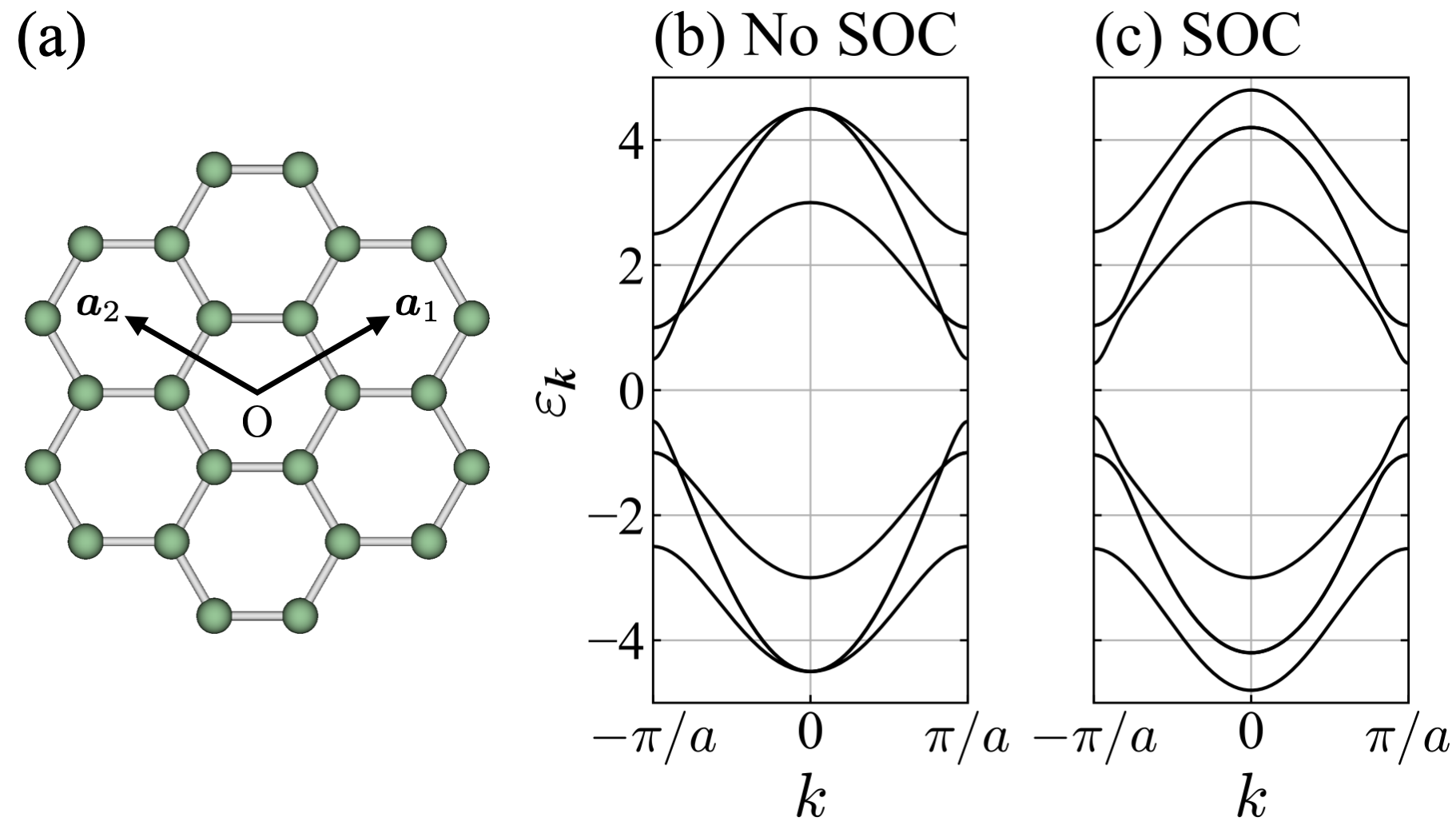}
        \caption{(a) Honeycomb lattice and its primitive translation vectors. (b)(c) Dispersion relations of the $p$ orbitals on a honeycomb lattice along $(k/\sqrt{3},k,0)^{\mathrm{T}}$ for the cases (b) without SOC and (c) with SOC. The Slater-Koster parameters are set as $V_{pp\sigma}=-2.0$ and $V_{pp\pi}=-1.0$. The SOC parameters for (c) are chosen as $K_{pp\sigma}=K_{pp\pi}'=0.1$.}
        \label{fig:disp_honeycomb}
    \end{center}
\end{figure}

Next, we consider the $p$ orbitals on a honeycomb lattice shown in Fig. \ref{fig:disp_honeycomb} (a).
Let $\bm{a}_1=(\sqrt{3}a/2,a/2,0)^{\mathrm{T}}$ and $\bm{a}_2=(-\sqrt{3}a/2,a/2,0)^{\mathrm{T}}$ be the primitive translation vectors.
Figure \ref{fig:disp_honeycomb} (b)(c) shows the band dispersion $\varepsilon_{\bm{k}}$ along $(k/\sqrt{3},k,0)^{\mathrm{T}}$ for the case of (b) without SOC and (c) with SOC.
The Slater-Koster parameters and the SOC parameters are set to the same as before.

\begin{figure}[b]
    \begin{center}
        \includegraphics[width=6cm]{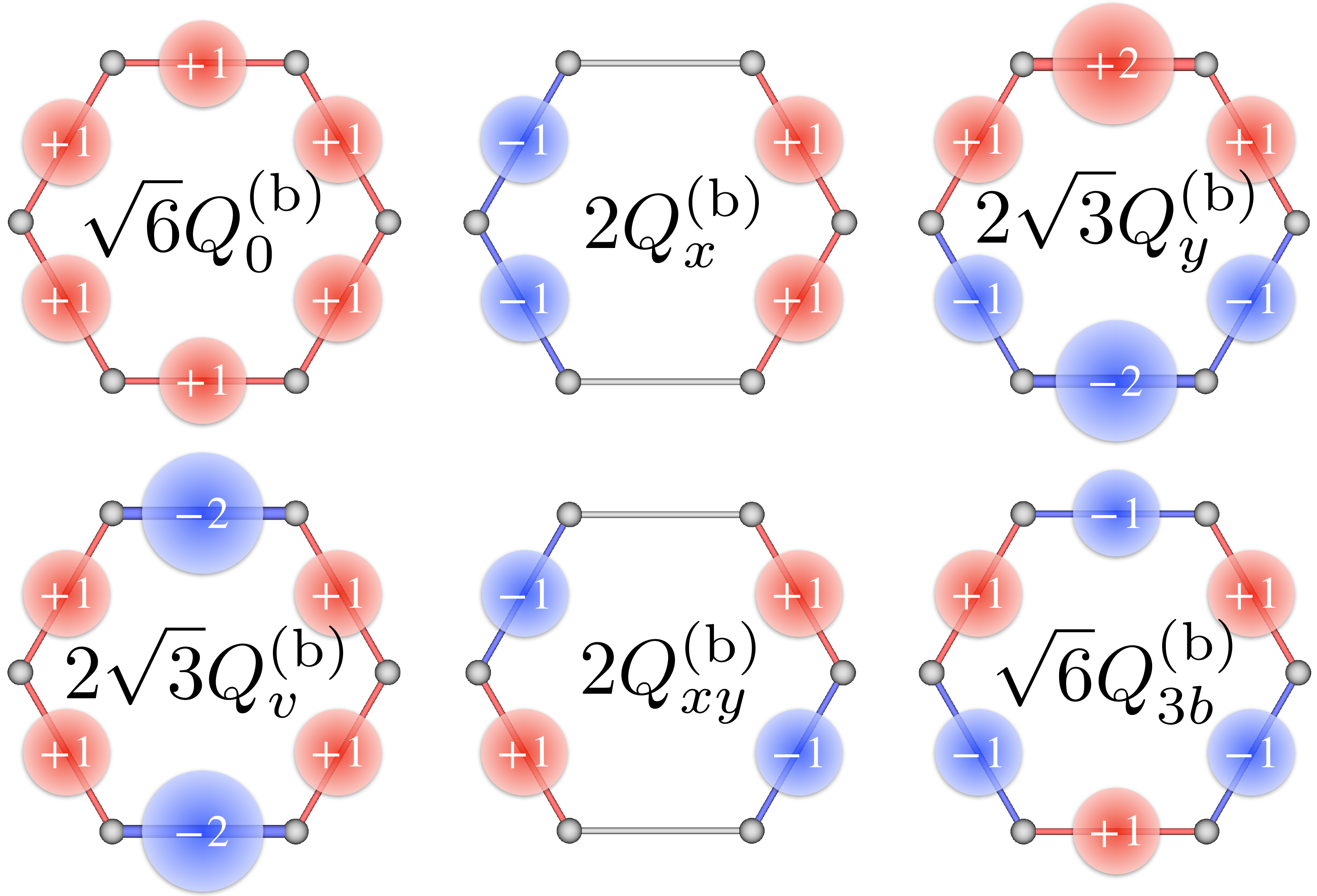}
        \caption{Bond multipoles for a hexagonal cluster. The red and blue spheres represent the presence of positive and negative charges at the bond centers, respectively. The gray bonds with no sphere represent the absence of a charge. The number in the spheres indicates the value of the charge on the bond. For a hexagonal cluster, there are six bond multipoles, each charge configuration corresponds to an electric multipole represented by $Q^{(\mathrm{b})}_X$, where $X=0,x,y,v,xy,3b$.}
        \label{fig:honeycomb_bond}
    \end{center}
\end{figure}

Similarly to the previous section, we conduct an analysis using multipoles to observe how the SOC affects the band structure.
For symmetry discussions, we consider a hexagonal cluster shown in Fig. \ref{fig:clusters} (b).
Figure \ref{fig:honeycomb_bond} shows the bond multipoles for the hexagonal cluster.
The cluster for the bond multipoles is composed of six bonds and there are six bond multipoles accordingly.
The number in the spheres indicates the value of the charge on the bond.

By adopting the same procedure as in the case of the square lattice, the multipole corresponding to the first term in Eq. (\ref{eq:ils}) is
\begin{align}
    \label{eq:honeycomb_multipoles1}
    \vcenter{\hbox{\includegraphics[width=1.2cm]{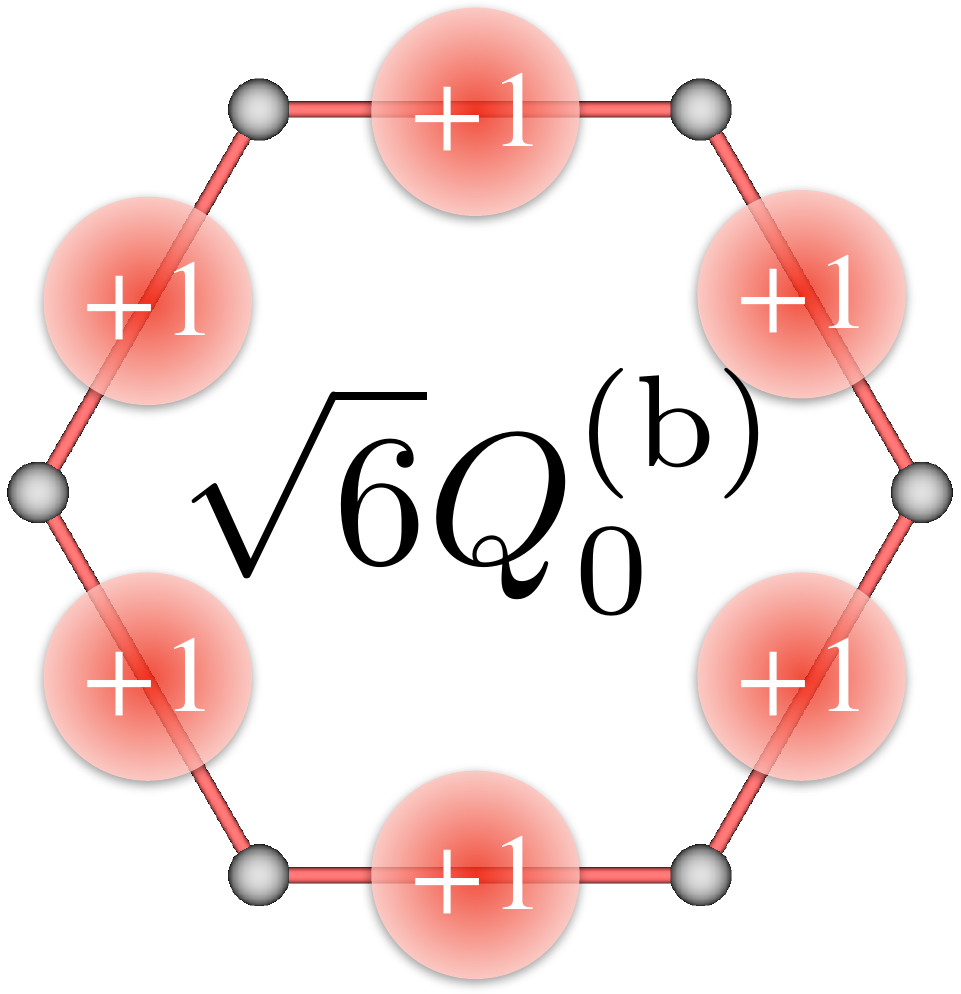}}}\otimes(\sum_{\nu=x,y,z}M_{\nu}^{(\mathrm{o})}\otimes M_{\nu}^{(\mathrm{s})}),
\end{align}
which leads to $Q_0^{(\mathrm{b})}\otimes Q_0^{(\mathrm{os})}=Q_0^{(\mathrm{bos})}$.
In the second term in Eq. (\ref{eq:ils}), $\hat{R}_{\nu}\hat{R}_{\nu'}$ depends on the bond.
For the bond in the $x$-direction, we have $\hat{\bm{R}}=\pm(1,0,0)^{\mathrm{T}}$ and in the other directions, $\hat{\bm{R}}=\pm(\frac{1}{2},\pm \frac{\sqrt{3}}{2},0)^{\mathrm{T}}$.
Thus, for the $x$-directional bond, we have only $\nu=\nu'=x$, i.e., $M_x^{(\mathrm{o})}\otimes M_x^{(\mathrm{s})}$.
Similarly, for the other bonds, we have 
\begin{align}
    & \frac{1}{4} M_x^{(\mathrm{o})}\otimes M_x^{(\mathrm{s})} + \frac{3}{4} M_y^{(\mathrm{o})}\otimes M_y^{(\mathrm{s})}  \nonumber \\
    & \pm \frac{\sqrt{3}}{4}\left (M_x^{(\mathrm{o})}\otimes M_y^{(\mathrm{s})} + M_y^{(\mathrm{o})}\otimes M_x^{(\mathrm{s})} \right ).
\end{align}

From these, we can see that the multipole corresponding to this second term is
\begin{align}
    \label{eq:honeycomb_multipoles2}
    &\frac{1}{2}\vcenter{\hbox{\includegraphics[width=1.2cm]{honeycomb_Q0_v3.pdf}}}\otimes(M_x^{(\mathrm{o})}\otimes M_x^{(\mathrm{s})}+M_y^{(\mathrm{o})}\otimes M_y^{(\mathrm{s})}) \nonumber \\
    &-\frac{1}{4}\vcenter{\hbox{\includegraphics[width=1.2cm]{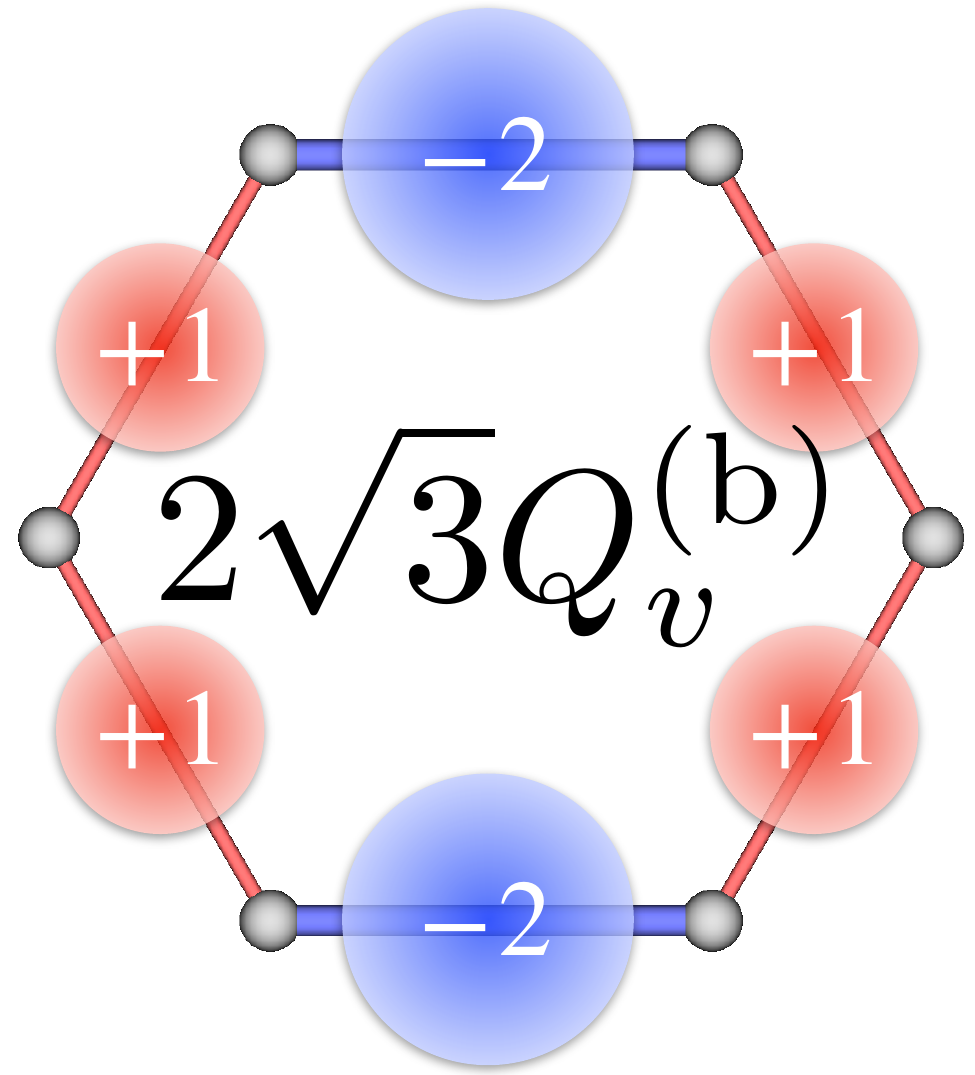}}}\otimes(M_x^{(\mathrm{o})}\otimes M_x^{(\mathrm{s})}-M_y^{(\mathrm{o})}\otimes M_y^{(\mathrm{s})}) \nonumber \\
    &-\frac{\sqrt{3}}{4}\vcenter{\hbox{\includegraphics[width=1.2cm]{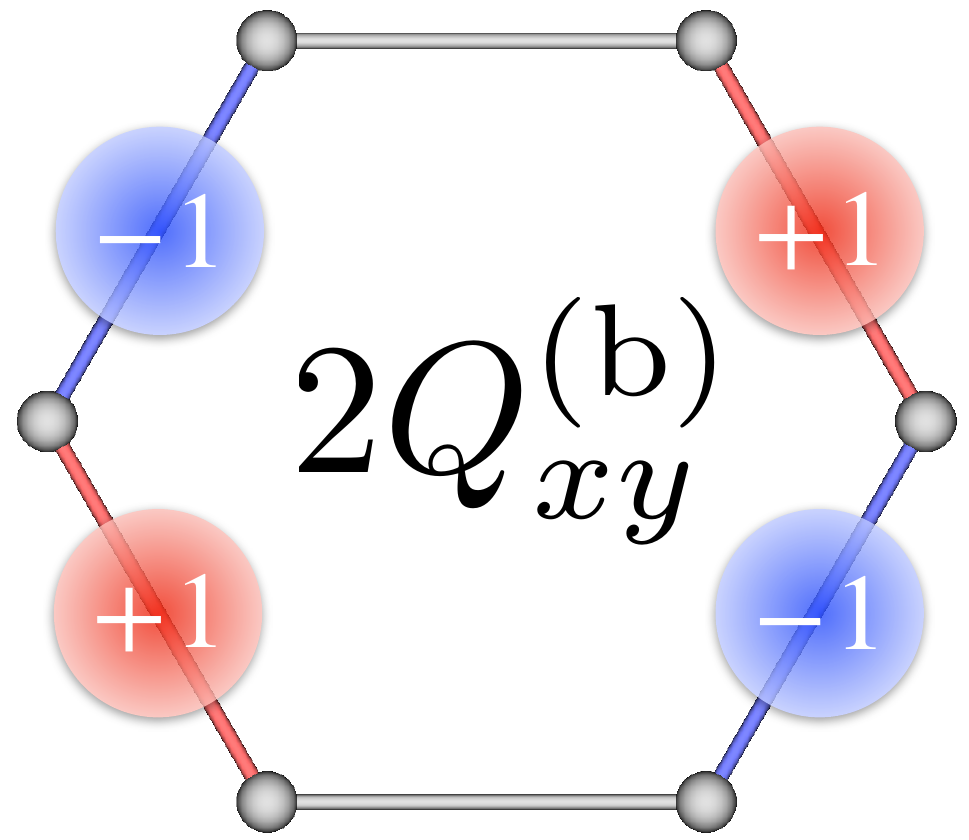}}}\otimes(M_x^{(\mathrm{o})}\otimes M_y^{(\mathrm{s})}+M_y^{(\mathrm{o})}\otimes M_x^{(\mathrm{s})}).
\end{align}

In the case of Eq. (\ref{eq:honeycomb_multipoles2}), the first term is proportional to
\begin{align}
    Q_0^{(\mathrm{b})}\otimes \left(\sqrt{3}Q_0^{(\mathrm{os})}-\frac{\sqrt{6}}{2}Q_u^{(\mathrm{os})}\right),
\end{align}
which contains $Q_0^{(\mathrm{bos})}$ and $Q_u^{(\mathrm{bos})}$.
The second and third term are written as 
\begin{align}
    Q_v^{(\mathrm{b})}\otimes \sqrt{2}Q_v^{(\mathrm{os})} + Q_{xy}^{(\mathrm{b})}\otimes \sqrt{2}Q_{xy}^{(\mathrm{os})},
\end{align}
which is decomposed as a sum of $Q_0^{(\mathrm{bos})}, Q_u^{(\mathrm{bos})}$ and  $Q_{40}^{(\mathrm{bos})}$.
Since all the multipoles appearing in $H_{\mathrm{SOC}}$ are even-rank electric multipoles, symmetric band deformation occur as discussed in the previous case.

\subsection{$p$ orbitals on a triangular helical chain (point group $D_3$)}
\begin{figure}[t]
    \begin{center}
        \includegraphics[width=5.5cm]{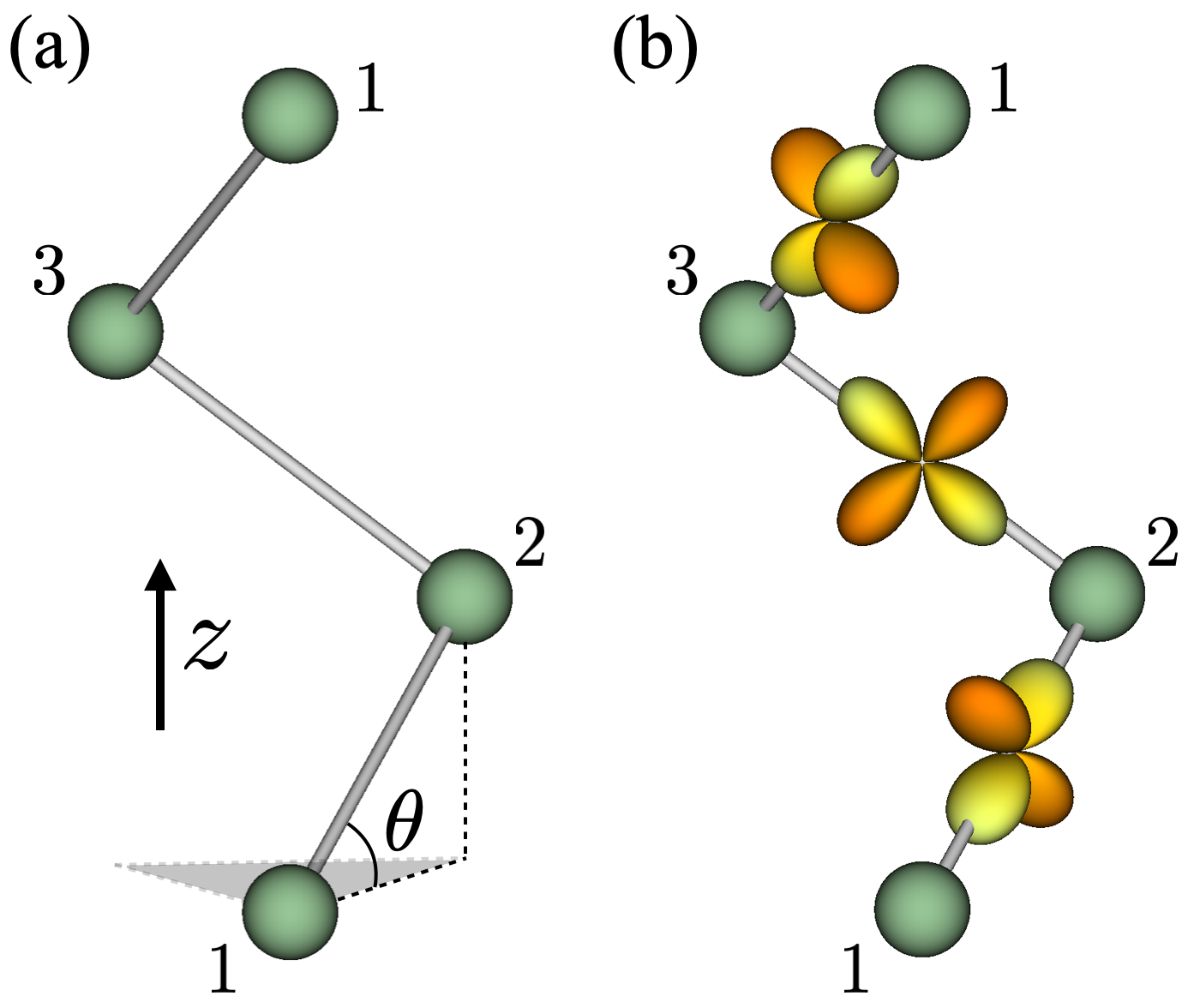}
        \caption{(a) Structure of the triangular helical chain ($1\to 2\to 3\to 1\to 2\to 3\to \cdots$). A parameter $\theta$ determines the angle of the bonds with respect to the $xy$-plane. (b) Schematic picture of the composite multipole $G_u^{(\mathrm{bos})}$, which corresponds to Eq. (\ref{eq:Gu_term}) in $H_{\mathrm{SOC}}$. The quadrupoles at the bond centers in yellow and orange represent the orbital-spin multipoles created by linear combinations of $Q_{yz}^{(\mathrm{os})}$ and $Q_{zx}^{(\mathrm{os})}$. Figures are created using QtDraw and MultiPie \cite{kusunose2023symmetry}.}
        \label{fig:trihelix_lattice}
    \end{center}
\end{figure}

Finally, we consider a one-dimensional triangular helical chain winding counterclockwise in the $z$-axis.
Figure \ref{fig:trihelix_lattice} (a) shows the structure of the triangular helical chain ($1\to 2\to 3\to 1\to 2\to 3\to \cdots$).
The length of the unit cell along the $z$-axis is set $a$, and $\theta$ is introduced as a degree of freedom to determine the pitch of the helix. 
$\theta$ is defined as the angle formed by each bond with the $xy$-plane ($0<\theta<\pi/2$) as shown in Fig. \ref{fig:trihelix_lattice} (a). Taking the limit of $\theta\to 0$, the pitch of the helix tends to zero, resulting in a triangle in the $xy$-plane.

\begin{figure}[b]
    \begin{center}
        \includegraphics[width=6.5cm]{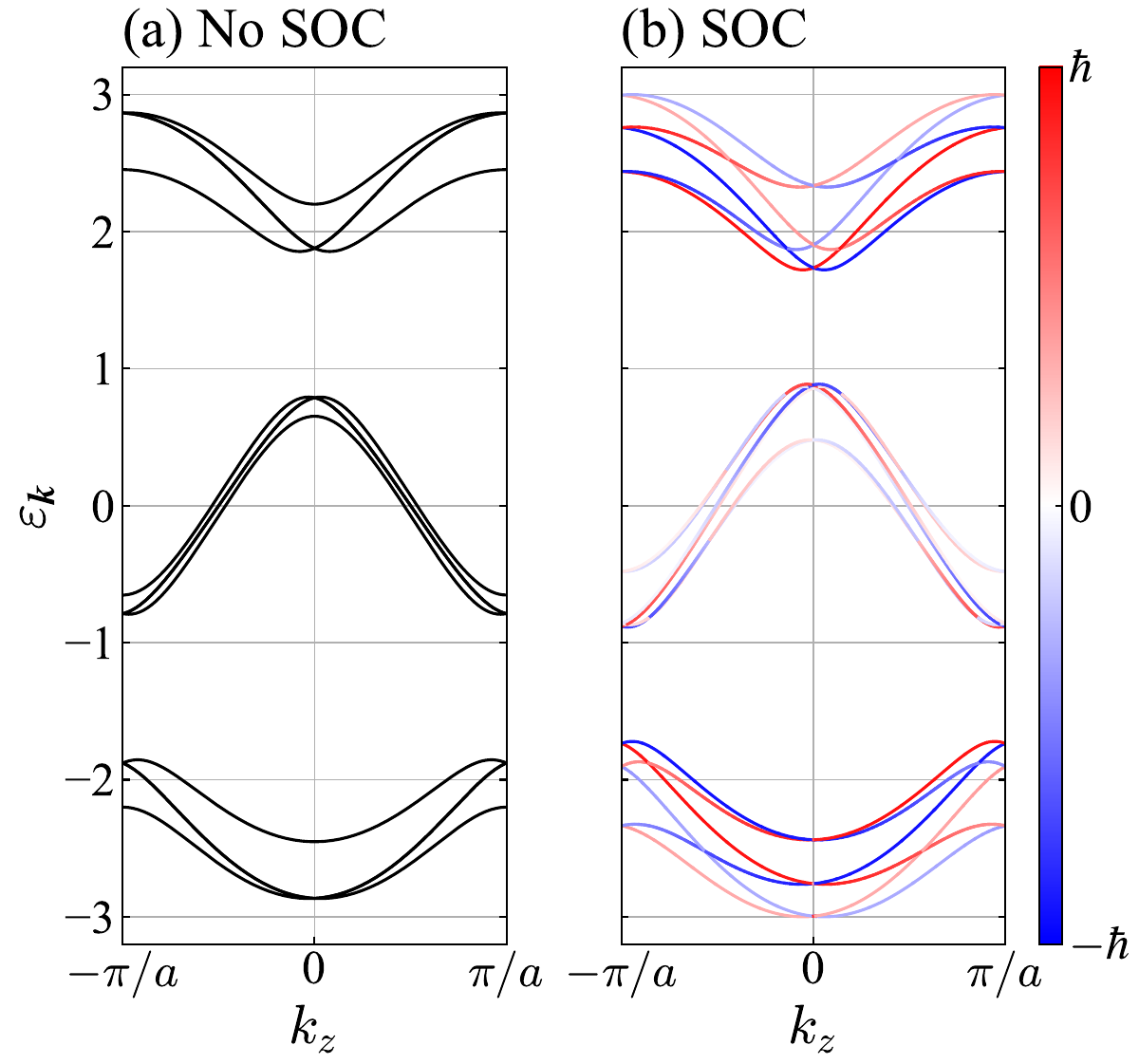}
        \caption{Dispersion relations of the $p$ orbitals on a triangular helical chain as a function of the wave vector $k_z$ for the cases (a) without SOC and (b) with SOC. The parameter angle $\theta$ is set such that $\tan\theta=1/2$. The Slater-Koster parameters are set as $V_{pp\sigma}=-2.0$ and $V_{pp\pi}=-1.0$. The SOC parameters for (b) are chosen as $K_{pp\sigma}=K_{pp\pi}'=0.1$.}
        \label{fig:disp_trihelix}
    \end{center}
\end{figure}

The dispersion relations (a) without SOC and (b) with SOC are shown in Fig. \ref{fig:disp_trihelix}. 
The Slater-Koster parameters and the SOC parameters are set to the same as before and $\theta$ is set such that $\tan\theta=1/2$.
The bands are colored based on the $z$-component spin polarization $\langle\sigma_z\rangle$.
In the previous cases in Figs. \ref{fig:disp_square} and \ref{fig:disp_honeycomb}, the up spin and the down spin are degenerate.
In contrast, in Fig. \ref{fig:disp_trihelix} (b), we can see that bands with opposite $\langle\sigma_z\rangle$ split around the $\Gamma$ point in an antisymmetric way with respect to $k_z$.
This characteristic antisymmetric spin splitting in the form of $k_z\sigma_z$ is unique to chiral systems \cite{yoda2015current, suzuki2023spin} and we refer to it as chiral spin splitting in the subsequent discussion.

\begin{figure}[t]
    \begin{center}
        \includegraphics[width=5.5cm]{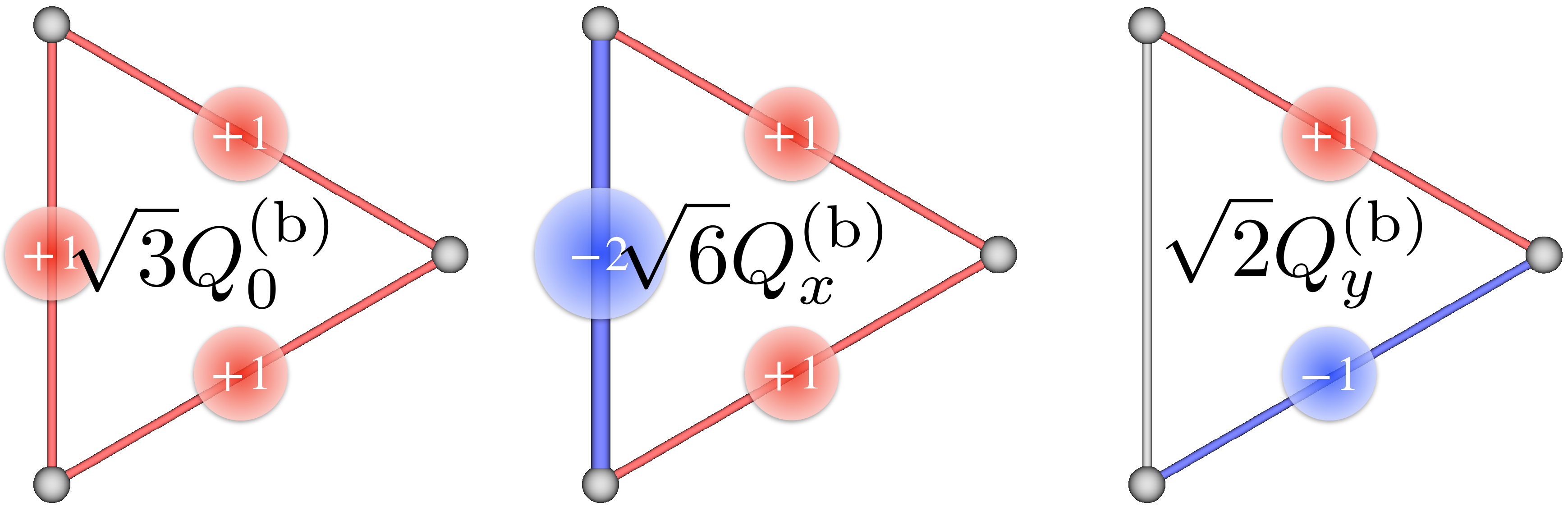}
        \caption{Bond multipoles for a triangular helical cluster. The figure shows the cluster viewed from the z-axis. The red and blue spheres represent the presence of positive and negative charges at the bond centers, respectively, which form the gray shaded triangle in Fig. \ref{fig:clusters} (c). The gray bond with no sphere on top represents the absence of a charge. The number in the spheres denotes the value of the charge. For a triangular cluster, there are three bond multipoles, each charge configuration corresponds to an electric multipole represented by $Q^{(\mathrm{b})}_X$, where $X=0,x,y$.}
        \label{fig:triangular_bond}
    \end{center}
\end{figure}

We analyze the chiral spin splitting from the perspective of multipoles.
For symmetry discussions, we consider a triangular helical cluster shown in Fig. \ref{fig:clusters} (c).
This cluster can be constructed by applying all symmetry operations of point group $D_3$ to one bond in the lattice.
The origin is chosen to be at the height of the bond center.
The bond multipoles for the triangular helical cluster are shown in Fig. \ref{fig:triangular_bond}.

Following the same method we used for the other lattices, the multipole corresponding to the first term in Eq. (\ref{eq:ils}) is
\begin{align}
    \label{eq:trihelix_multipoles1}
    \vcenter{\hbox{\includegraphics[width=1.2cm]{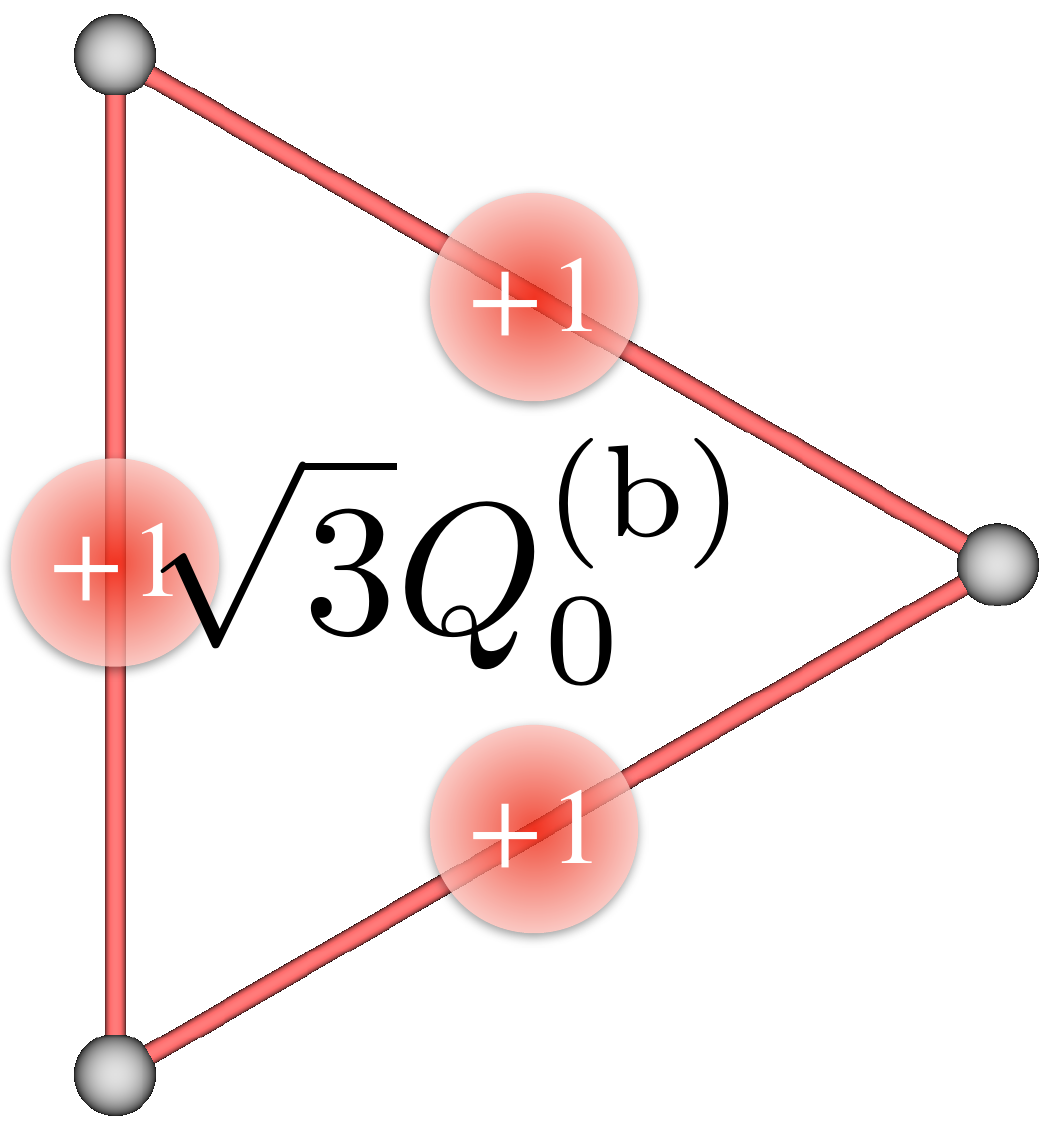}}}\otimes(\sum_{\nu=x,y,z}M_{\nu}^{(\mathrm{o})}\otimes M_{\nu}^{(\mathrm{s})}),
\end{align}
which leads to $Q_0^{(\mathrm{b})}\otimes Q_0^{(\mathrm{os})}=Q_0^{(\mathrm{bos})}$.

The second term in Eq. (\ref{eq:ils}) depends on the bond through $\hat{R}_{\nu}\hat{R}_{\nu'}$.
For the three bonds, we have 
\begin{align}
    \label{eq:R_trihelix}
    \hat{\bm{R}}=
    \pm\begin{pmatrix}
        0 \\
        -\cos\theta \\
        \sin\theta
    \end{pmatrix},
    \pm\begin{pmatrix}
        \frac{\sqrt{3}}{2}\cos\theta \\
        \frac{1}{2}\cos\theta \\
        \sin\theta
    \end{pmatrix},
    \pm\begin{pmatrix}
        -\frac{\sqrt{3}}{2}\cos\theta \\
        \frac{1}{2}\cos\theta \\
        \sin\theta
    \end{pmatrix},
\end{align}
which result in 
\begin{align}
    & \cos^2\theta M_y^{(\mathrm{o})}\otimes M_y^{(\mathrm{s})} + \sin^2\theta M_z^{(\mathrm{o})}\otimes M_z^{(\mathrm{s})}  \nonumber \\
    & -\sin\theta\cos\theta \left ( M_y^{(\mathrm{o})}\otimes M_z^{(\mathrm{s})} + M_z^{(\mathrm{o})}\otimes M_y^{(\mathrm{s})} \right ), \\
    & \cos^2\theta \left(\frac{3}{4}M_x^{(\mathrm{o})}\otimes M_x^{(\mathrm{s})} + \frac{1}{4} M_y^{(\mathrm{o})}\otimes M_y^{(\mathrm{s})} \right)  \nonumber \\
    & + \sin^2\theta M_z^{(\mathrm{o})}\otimes M_z^{(\mathrm{s})} \nonumber \\
    & \pm \frac{\sqrt{3}}{4}\cos^2\theta \left(M_x^{(\mathrm{o})}\otimes M_y^{(\mathrm{s})} + M_y^{(\mathrm{o})}\otimes M_x^{(\mathrm{s})}\right)\nonumber \\
    & + \frac{1}{2}\sin\theta \cos\theta \left(M_y^{(\mathrm{o})}\otimes M_z^{(\mathrm{s})} + M_z^{(\mathrm{o})}\otimes M_y^{(\mathrm{s})}\right)\nonumber \\
    & \pm \frac{\sqrt{3}}{2}\sin\theta \cos\theta \left(M_z^{(\mathrm{o})}\otimes M_x^{(\mathrm{s})} + M_x^{(\mathrm{o})}\otimes M_z^{(\mathrm{s})}\right), \nonumber \\
\end{align}
for each bond, where the double signs are in the same order.
From these, we can see that the multipole corresponding to this second term is
\begin{align}
    \label{eq:trihelix_multipoles2-1}
    &\left(\frac{1}{2}\cos^2\theta\vcenter{\hbox{\includegraphics[width=1.2cm]{triangular_Q0_v3.pdf}}}\otimes(M_x^{(\mathrm{o})}\otimes M_x^{(\mathrm{s})}+M_y^{(\mathrm{o})}\otimes M_y^{(\mathrm{s})})\right. \nonumber \\
    &\left.+\sin^2\theta\vcenter{\hbox{\includegraphics[width=1.2cm]{triangular_Q0_v3.pdf}}}\otimes M_z^{(\mathrm{o})}\otimes M_z^{(\mathrm{s})}\right) \\
    \label{eq:trihelix_multipoles2-2}
    &+\frac{1}{2}\cos^2\theta\left(\frac{1}{2}\vcenter{\hbox{\includegraphics[width=1.2cm]{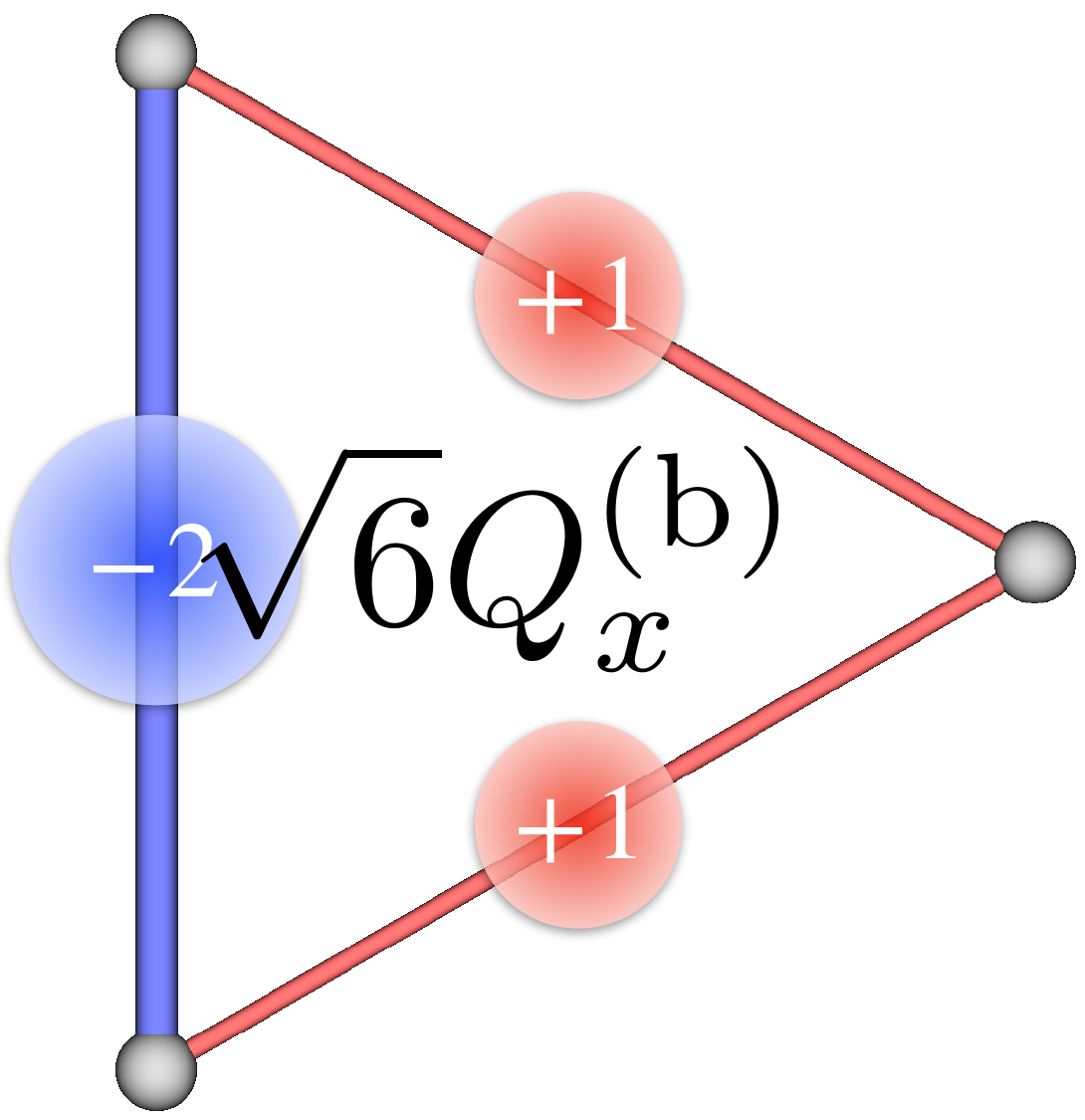}}}\otimes(M_x^{(\mathrm{o})}\otimes M_x^{(\mathrm{s})}-M_y^{(\mathrm{o})}\otimes M_y^{(\mathrm{s})})\right. \nonumber \\
    &\left.-\frac{\sqrt{3}}{2}\vcenter{\hbox{\includegraphics[width=1.2cm]{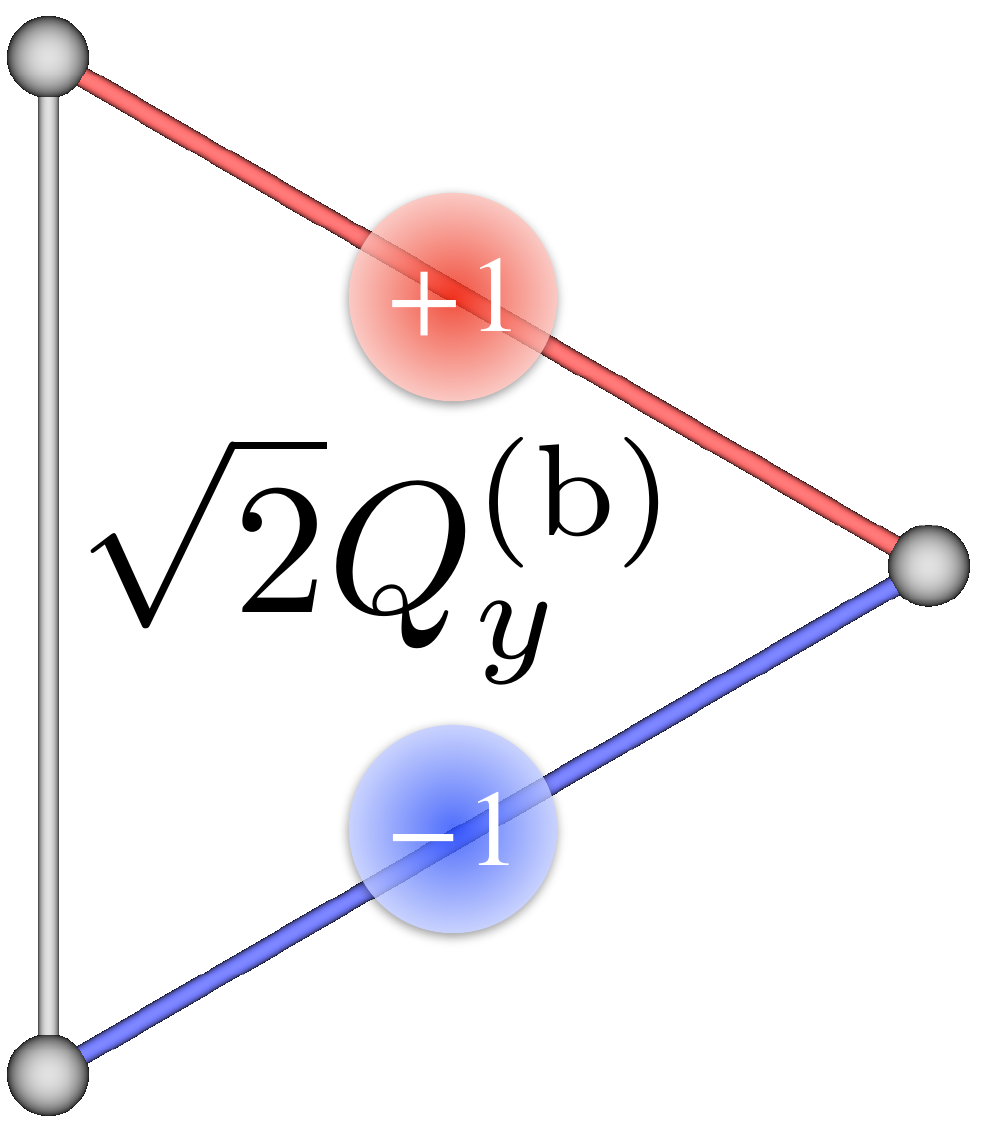}}}\otimes(M_x^{(\mathrm{o})}\otimes M_y^{(\mathrm{s})}+M_y^{(\mathrm{o})}\otimes M_x^{(\mathrm{s})})\right) \\
    \label{eq:trihelix_multipoles2-3}
    &+\frac{1}{2}\sin 2\theta\left(\frac{1}{2}\vcenter{\hbox{\includegraphics[width=1.2cm]{triangular_Qx_v3.pdf}}}\otimes(M_y^{(\mathrm{o})}\otimes M_z^{(\mathrm{s})}+M_z^{(\mathrm{o})}\otimes M_y^{(\mathrm{s})})\right. \nonumber \\
    &\left.-\frac{\sqrt{3}}{2}\vcenter{\hbox{\includegraphics[width=1.2cm]{triangular_Qy_v3.pdf}}}\otimes(M_z^{(\mathrm{o})}\otimes M_x^{(\mathrm{s})}+M_x^{(\mathrm{o})}\otimes M_z^{(\mathrm{s})})\right).
\end{align}

In the case of Eq. (\ref{eq:trihelix_multipoles2-1}), we have
\begin{align}
    Q_0^{(\mathrm{b})}\otimes \left( Q_0^{(\mathrm{os})} + \frac{1}{\sqrt{2}} (3\sin^2\theta-1)Q_u^{(\mathrm{os})} \right),
\end{align}
which gives $Q_0^{(\mathrm{bos})}$ and $Q_u^{(\mathrm{bos})}$.
In the case of Eq. (\ref{eq:trihelix_multipoles2-2}), we have
\begin{align}
    \frac{\sqrt{3}}{2}\cos^2\theta(Q_x^{(\mathrm{b})}\otimes Q_v^{(\mathrm{os})}-Q_y^{(\mathrm{b})}\otimes Q_{xy}^{(\mathrm{os})}),
\end{align}
which gives $Q_{3a}^{(\mathrm{bos})}$.
In the case of Eq. (\ref{eq:trihelix_multipoles2-3}), we have
\begin{align}
    \label{eq:Gu_term}
    \frac{\sqrt{3}}{2}\sin 2\theta(Q_x^{(\mathrm{b})}\otimes Q_{yz}^{(\mathrm{os})}-Q_y^{(\mathrm{b})}\otimes Q_{zx}^{(\mathrm{os})}),
\end{align}
which gives $G_u^{(\mathrm{bos})}$.
Here, $G_u^{(\mathrm{bos})}$ is an electric toroidal quadrupole characteristic to chiral systems \cite{kishine2022definition, hirose2022electronic, tsunetsugu2023theory}.

Since the triangular helical chain breaks inversion symmetry, odd-rank and even-rank multipoles cannot be distinguished in a sense that they may belong to the same irreducible representation (irrep) \cite{hayami2020bottom}.
For example, the bond electric monopole $Q_0^{(\mathrm{b})}$ could be named as an electric octupole $Q_{3a}^{(\mathrm{b})}$, since both of them belong to the $A_1$ representation.
Therefore, the presence of an odd-rank multipole in the SOC Hamiltonian cannot be attributed as the cause of antisymmetric spin splitting in the band dispersion.

Despite this ambiguity, we can identify the source of the chiral spin splitting in Fig. \ref{fig:disp_trihelix} (b) as follows.
Noting that the set of the three bond centers of the triangular helical cluster remains invariant under all symmetry operations of the $D_{3h}$ point group, the bond multipoles for the triangular helical chain belonging to $D_3$ can be viewed as irreps of the supergroup $D_{3h}$.
Thus, we find that the $G_u^{(\mathrm{bos})}$ term (Eq. (\ref{eq:Gu_term})) belongs to the $A_1''$ irrep in $D_{3h}$ by considering additional symmetry operations in $D_{3h}$.  
Since the $A_1''$ irrep breaks the mirror symmetry with respect to the $xy$-plane, the band dispersion $\varepsilon_{n\bm{k}\sigma}$ along the $\Gamma$-$A$ line ($\bm{k}=(0,0,k_z)$) can be unequal to $\varepsilon_{n''-\bm{k}\sigma}$, where $n$ and $n''$ are band indices that transform into each other under mirror reflection.
This indicates that $\varepsilon_{n\bm{k}\sigma}$ is not equal to $\varepsilon_{\overline{n}''\bm{k}-\sigma}$ combined with Eq. (\ref{eq:TRS}), resulting in the chiral band splitting.
The coefficient $\sin2\theta$ in Eq. (\ref{eq:trihelix_multipoles2-3}) also shows that the $G_u^{(\mathrm{bos})}$ term vanishes if the system falls into $D_{3h}$ with $\theta\to 0$.

In Fig. \ref{fig:trihelix_lattice} (b), the schematic picture of $G_u^{(\mathrm{bos})}$ is shown.
The quadrupoles at the bond centers in yellow and orange represent the orbital-spin multipoles created by linear combinations of $Q_{yz}^{(\mathrm{os})}$ and $Q_{zx}^{(\mathrm{os})}$.
We can see that the orbital-spin quadrupoles are aligned along the helix, giving the chiral nature to the SOC Hamiltonian.

We note that the chiral band splitting occurs without the $G_u^{(\mathrm{bos})}$ term of Eq. (\ref{eq:trihelix_multipoles2-3}), because the spin-independent hopping also gives rise to a term that breaks the mirror symmetry.
The second term of the spin-independent hopping integral in Eq. (\ref{eq:t_parameter}), which depends on the bond angle, becomes $Q_{3a}^{(\mathrm{bos})}$ and $G_u^{(\mathrm{bos})}$ by performing multipole decomposition of the corresponding Hamiltonian.
The details of the multipole decomposition are discussed in Appendix \ref{app:spin-independent_multipoles}.
Therefore, if there is any SOC Hamiltonian that involves the $z$-component of the spin, it should cause the chiral spin splitting.
This can be demonstrated by setting $K_{pp\sigma}=-K_{pp\pi}'=0.1$, which removes Eqs. (\ref{eq:trihelix_multipoles2-1})-(\ref{eq:trihelix_multipoles2-3}).
In this case, we still observe the chiral spin splitting.
But the chiral spin splitting does not occur if we further tune to $V_{pp\sigma}=V_{pp\pi}=-2$, which gets rid of the bond dependence of $H_{\mathrm{kin}}$.

Importantly, we find that an electric toroidal monopole $G_0^{(\mathrm{bos})}$ is not present in the obtained SOC Hamiltonian although it is symmetrically allowed to appear.
We can prove that $G_0^{(\mathrm{bos})}$ is not generated within the two-center approximation as far as the $p$ orbitals are concerned as follows.
First, note that the bond multipoles are all electric multipoles and that the orbital-spin composite multipoles are an electric monopole, electric quadrupoles or electric toroidal dipoles as listed in Eqs. (\ref{eq:os_basis1})-(\ref{eq:os_basis9}), it is necessary to combine the orbital-spin composite multipoles $G_x^{(\mathrm{os})}$ and $G_y^{(\mathrm{os})}$ with the bond multipoles $Q_x^{(\mathrm{b})}$ and $Q_y^{(\mathrm{b})}$ in order to obtain $G_0^{(\mathrm{bos})}$.
However, the SOC hopping integral in Eq. (\ref{eq:ils}) has only orbital-spin symmetric terms, which cannot contain $G_x^{(\mathrm{os})}$ or $G_y^{(\mathrm{os})}$.
Therefore, $G_0^{(\mathrm{bos})}$ is not generated.

\section{Summary\label{sec:summary}}
In summary, we have established the microscopic foundation of the SOC hopping within the two-center approximation.
Starting with the general form of the SOC, we have analytically derived the explicit expression for the spin-dependent hopping integrals applicable to any atomic orbitals and hopping directions, which are expressed as extended Slater-Koster symbols. 
Using the derived SOC hopping, we have investigated how it affects the band structure based on tight-binding models on several lattices.
By carrying out a multipole decomposition for the Hamiltonians, we have demonstrated that the SOC hoppings give rise to band deformation or spin splitting depending on the symmetry of the lattice. 
Our formulation of the spin-dependent hopping based on atomic orbitals and its classification by multipoles offer a valuable framework for constructing microscopic models for spin transport phenomena in specific materials, especially when addressing unique phenomena originating from the symmetry of the system such as the spin Hall effect and the CISS effect.

In the case of the chiral triangular helical chain, we have attributed the cause of the chiral spin splitting to an electric toroidal quadrupole $G_u^{(\mathrm{bos})}$.
Notably, the absence of the lowest-rank electric toroidal monopole $G_0^{(\mathrm{bos})}$ within the two-center approximation, despite that it is symmetrically allowed, suggests the necessity of exloring more intricate schemes such as incorporating a third center or hybridizing orbitals with different parities.
This observation underscores the potential need to consider higher-rank electric toroidal multipoles beyond $G_0$ to quantify chirality, given that the expectation value of $G_0$ might be small in some cases.
Along with recent attempts to quantify chirality \cite{oiwa2022rotation, hayami2023chiral, Inda_2024}, these insights provide guidance for future strategies in defining an appropriate chirality measure.

\begin{acknowledgments}
    We thank Hiroaki Kusunose, Rikuto Oiwa and Jun-ichiro Kishine for fruitful discussions.
    This work was supported by JSPS KAKENHI Grant No. JP23K03274, MEXT, Japan.
    Masaki Kato was supported by World-leading Innovative Graduate Study Program for Materials Research, Industry, and Technology (MERIT-WINGS).
\end{acknowledgments}

\appendix
\begin{widetext}
\section{Wigner D-matrix}
\label{app:WDM}
Let $R(\alpha,\beta,\gamma)=e^{-i\alpha J_z}e^{-i\beta J_y}e^{-i\gamma J_z}$ be an rotation operator, where $\alpha, \beta, \gamma$ are Euler angles and $J_{\mathrm{\mu}}$ is the angular momentum operator about the $\mu(=x,y,z)$ axis, respectively.
Let $|jm\rangle$ be a simultaneous eigenstate of $J^2=\sum_{\mu=x,y,z}J_{\mu}^2$ and $J_z$ that satisfies 
\begin{align}
    J^2|jm\rangle&=j(j+1)|jm\rangle, \\
    J_z|jm\rangle&=m|jm\rangle.
\end{align}
The Wigner D-matrix $D_{m'm}^{(j)}(\alpha,\beta,\gamma)$ is defined as the matrix element of $R(\alpha,\beta,\gamma)$ with respect to $\langle jm'|$ and $|jm\rangle$ \cite{varshalovich1988quantum}:
\begin{align}
    D_{m'm}^{(j)}(\alpha,\beta,\gamma):=\langle jm'|R(\alpha,\beta,\gamma)|jm\rangle.
\end{align}
$D_{m'm}^{(j)}(\alpha,\beta,\gamma)$ can be written as 
\begin{align}
    D_{m'm}^{(j)}(\alpha,\beta,\gamma) = e^{-im'\alpha}d_{m'm}^j(\beta)e^{-im\gamma},
\end{align}
where 
\begin{align}
    d_{m'm}^j(\beta)=(-1)^{m'-m}[(j+m')!(j-m')!(j+m)!(j-m)!]^{\frac{1}{2}}\times\sum_k\left[\frac{(-1)^k(\cos\frac{\beta}{2})^{2j+m-m'-2k}(\sin\frac{\beta}{2})^{m'-m+2k}}{k!(j+m-k)!(m'-m+k)!(j-m'-k)!}\right].
\end{align}

\section{The definition of the bond multipoles}
\label{app:bondmultipoles}
Let us denote the set of bonds obtained by applying all the symmetry operations of a point group to a single bond as a cluster.
If we assume placing electric charges at the bond centers, we can find symmetry-adapted charge configurations that belong to an irreducible representation (irrep) of the point group.
The number of the independent charge configurations should be the number of bonds $N_{\mathrm{bond}}$.

Recalling that for an $N$-electron sysmtem with elementary charge $-e$, the electric multipoles are given by
\begin{align}
    Q_{lm} = -e\sum_{i=1}^N O_{lm}(\bm{r}_i),
\end{align}
where $\bm{r}_i$ represents the position of the $i$-th electron and $O_{lm}(\bm{r}) = \sqrt{\frac{4\pi}{2l+1}}r^lY_{lm}(\hat{\bm{r}})$, electric multipoles representing the charge configuration that belongs to an irrep $\Gamma$ can be defined by
\begin{align}
    Q_{lm}^{(\mathrm{b})} = \sum_{(ij)}^{N_{\mathrm{bond}}}q^{(ij)}_{\Gamma}O_{lm}(\bm{R}_{(ij)}),
\end{align}
where $(ij)$ represents the bond connecting the $i$-th and $j$-th sites, $\bm{R}_{(ij)}$ is the position of the center of $(ij)$ and $q^{(ij)}_{\Gamma}$ is a charge at $\bm{R}_{ij}$ in the irrep $\Gamma$.
The charge configuration at the bond centers $\{q^{(ij)}_{\Gamma}|(ij)\}$ is referred to as a bond multipole.
We note that more than one multipoles with different names could correspond to a given charge configuration.
In this case, we adopt only the one with the lowest rank to designate the charge configuration.
With the use of this notation, we rewrite $q^{(ij)}_{\Gamma}$ as $q^{(ij)}_{lm}$

The matrix form of the bond multipoles with respect to the atomic orbital bases is obtained by 
\begin{align}
    \left[\hat{Q}_{lm}^{(\mathrm{b})}\right]_{ij} = \left[\hat{Q}_{lm}^{(\mathrm{b})}\right]_{ji} = q^{(ij)}_{lm},
\end{align}
where a hat is used to denote the bond multipole as the matrix form.

For practical use of the bond multipoles, a particular normalization is used in the main text such that
\begin{align}
    \frac{1}{2}\mathrm{Tr}\left[\hat{Q}_{lm}^{(\mathrm{b})}\hat{Q}_{l'm'}^{(\mathrm{b})}\right] = \sum_{(ij)}q^{(ij)}_{lm}q^{(ij)}_{l'm'} \equiv \delta_{ll'}\delta_{mm'}.
\end{align}

\section{Correspondence between electric multipoles and tesseral, cubic and hexagonal harmonics}
\label{app:convention}
It is convenient to use real-valued tesseral harminics $\mathcal{Y}_{lm}$ to express multipoles rather than using spherical harmonics $Y_{lm}$:
\begin{align}
    \label{eq:O_harmtotess}
    \mathcal{O}_{l\mu}(\bm{r}) = \sum_{m=-l}^l C_{\mu m}O_{lm}(\bm{r}),
\end{align}
where $C_{\mu m}$ is a coefficient to transform the spherical harmonics to the tesseral harmonics defined in Eq. (\ref{eq:harmtotess}).
We note that the definition of $O_{lm}$ may differ in literatures.
When another definition $O_{lm}(\bm{r}) = \sqrt{\frac{4\pi}{2l+1}}r^lY^*_{lm}(\hat{\bm{r}})$ is used, Eq. (\ref{eq:O_harmtotess}) needs to be modified to $\mathcal{O}_{l\mu}(\bm{r}) = \sum_{m=-l}^l C^*_{\mu m}O_{lm}(\bm{r})$.

Conventionally, particular linear combinations of tesseral harmonics known as cubic harmonics and hexagonal harmonics are often used to represent multipoles in a form compatible to the symmetry of cubic or hexagonal point groups \cite{hayami2020bottom}.
We summarize them in the following table. 
The names of the cubic and hexagonal harmonics are adoptable to all types of multipoles including the electric toroidal, magnetic and magnetic toroidal multipoles, since they are defined through $O_{lm}$.

\begin{table}[H]
    \renewcommand{\arraystretch}{1.4}
    \centering
    \begin{adjustbox}{valign=t}
        \begin{minipage}[t]{0.43\linewidth}
            \centering
            \begin{tabular}{ccc} \hline
                \multicolumn{3}{c}{Cubic harmonics} \\ \hline
                Rank & Symbol & Correspondence \\ \hline
                0 & $Q_0$ & $(00)$ \\ \hline
                1 & $Q_x, Q_y, Q_z$ & $(11), (1\overline{1}), (10)$ \\ \hline
                2 & $Q_u, Q_v$ & $(20), (22)$ \\
                & $Q_{zx}, Q_{yz}, Q_{xy}$ & $(21), (2\overline{1}), (2\overline{2})$ \\ \hline
                3 & $Q_{xyz}$ & $(3\overline{2})$ \\
                & $Q_x^{\alpha}$ & $\frac{1}{2\sqrt{2}}[\sqrt{5}(33)-\sqrt{3}(31)]$ \\
                & $Q_y^{\alpha}$ & $-\frac{1}{2\sqrt{2}}[\sqrt{5}(3\overline{3})+\sqrt{3}(3\overline{1})]$ \\
                & $Q_z^{\alpha}$ & $(30)$ \\
                & $Q_x^{\beta}$ & $-\frac{1}{2\sqrt{2}}[\sqrt{3}(33)+\sqrt{5}(31)]$ \\
                & $Q_y^{\beta}$ & $\frac{1}{2\sqrt{2}}[-\sqrt{3}(3\overline{3})+\sqrt{5}(3\overline{1})]$ \\
                & $Q_z^{\beta}$ & $(32)$ \\ \hline
                4 & $Q_4$ & $\frac{1}{2\sqrt{3}}[\sqrt{5}(44)+\sqrt{7}(40)]$ \\
                & $Q_{4u}$ & $-\frac{1}{2\sqrt{3}}[\sqrt{7}(44)-\sqrt{5}(40)]$ \\
                & $Q_{4v}$ & $-(42)$ \\
                & $Q_{4x}^{\alpha}$ & $-\frac{1}{2\sqrt{2}}[(4\overline{3})+\sqrt{7}(4\overline{1})]$ \\
                & $Q_{4y}^{\alpha}$ & $-\frac{1}{2\sqrt{2}}[(43)-\sqrt{7}(41)]$ \\
                & $Q_{4z}^{\alpha}$ & $(4\overline{4})$ \\
                & $Q_{4x}^{\beta}$ & $\frac{1}{2\sqrt{2}}[\sqrt{7}(4\overline{3})-(4\overline{1})]$ \\
                & $Q_{4y}^{\beta}$ & $-\frac{1}{2\sqrt{2}}[\sqrt{7}(43)+(41)]$ \\
                & $Q_{4z}^{\beta}$ & $(4\overline{2})$ \\ \hline
            \end{tabular}
        \end{minipage}
    \end{adjustbox}
    \begin{adjustbox}{valign=t}
        \begin{minipage}[t]{0.43\linewidth}
            \centering
            \begin{tabular}{ccc} \hline
                \multicolumn{3}{c}{Hexagonal harmonics} \\ \hline
                Rank & Symbol & Correspondence \\ \hline
                0 & $Q_0$ & $(00)$ \\ \hline
                1 & $Q_z$ & $(10)$ \\
                    & $Q_x, Q_y$ & $(11), (1\overline{1})$ \\ \hline
                2 & $Q_u$ & $(20)$ \\
                    & $Q_v, Q_{xy}$ & $(22), (2\overline{2})$ \\
                    & $Q_{zx}, Q_{yz}$ & $(21), (2\overline{1})$ \\ \hline
                3 & $Q_z^{\alpha}$ & $(30)$ \\
                    & $Q_{3a}$ & $(33)$ \\
                    & $Q_{3b}$ & $(3\overline{3})$ \\
                    & $Q_{3u}, Q_{3v}$ & $(31), (3\overline{1})$ \\
                    & $Q_{z}^{\beta}, Q_{xyz}$ & $(32), (3\overline{2})$ \\ \hline
                4 & $Q_{40}$ & $(40)$ \\
                    & $Q_{4a}$ & $(4\overline{3})$ \\
                    & $Q_{4b}$ & $(43)$ \\
                    & $Q_{4u}^{\alpha}, Q_{4v}^{\alpha}$ & $(41), (4\overline{1})$ \\
                    & $Q_{4u}^{\beta 1}, Q_{4v}^{\beta 1}$ & $(44), (4\overline{4})$ \\
                    & $Q_{4u}^{\beta 2}, Q_{4v}^{\beta 2}$ & $(42), (4\overline{2})$ \\ \hline
            \end{tabular}
        \end{minipage}
    \end{adjustbox}
    \caption{Tables for correspondence between the electric harmonics in the form of the tesseral harmonics and (Left) the cubic harmonics and (Right) the hexagonal harmonics up to rank 4. The notations of $(l\mu)=\mathcal{O}_{l\mu}$ and $(l\overline{\mu})=\mathcal{O}_{l-\mu}$ are used for simplicity.}
\end{table}

\section{Multipole decomposition of spin-independent hopping Hamiltonian on a triangular helical chain}
\label{app:spin-independent_multipoles}
Spin-independent hopping Hamiltonian can also be decomposed into multipoles by using the same method used for the SOC Hamiltonian.
In this appendix, we discuss spin-independent hopping on a triangular helical chain.
Since the spin-independent hopping integral for $p$ orbitals $t_{\mu;\mu'}$ (Eq. (\ref{eq:t_parameter})) is symmetric with respect to the interchange of the orbitals, it can be expressed by the sum of the following symmteric matrices:
\begin{align}
    & Q_0^{(\mathrm{o})}=\frac{1}{\sqrt{3}}
    \begin{pmatrix}
        1 & & \\
        & 1 & \\
        & & 1
    \end{pmatrix}, 
    Q_u^{(\mathrm{o})}=\frac{1}{\sqrt{6}}
    \begin{pmatrix}
        -1 & & \\
        & -1 & \\
        & & 2
    \end{pmatrix},
    Q_v^{(\mathrm{o})}=\frac{1}{\sqrt{2}}
    \begin{pmatrix}
        1 & & \\
        & -1 & \\
        & & 0
    \end{pmatrix}, \nonumber \\
    & Q_{xy}^{(\mathrm{o})}=\frac{1}{\sqrt{2}}
    \begin{pmatrix}
        & 1 & \\
        1 & & \\
        & & 0
    \end{pmatrix},
    Q_{yz}^{(\mathrm{o})}=\frac{1}{\sqrt{2}}
    \begin{pmatrix}
        0 & & \\
        & & 1 \\
        & 1 & 
    \end{pmatrix},
    Q_{zx}^{(\mathrm{o})}=\frac{1}{\sqrt{2}}
    \begin{pmatrix}
        & & 1 \\
        & 0 & \\
        1 & & 
    \end{pmatrix},
\end{align}
where each column and row represent $p_x, p_y$ and $p_z$ orbitals.

Noting that the first term in Eq. (\ref{eq:t_parameter}) does not depend on the bond angle and is diagonal with respect to both orbital and spin, it becomes
\begin{align}
    \label{eq:spin-independent_multipoles1}
    \vcenter{\hbox{\includegraphics[width=1.2cm]{triangular_Q0_v3.pdf}}}\otimes Q_0^{(\mathrm{o})}\otimes Q_0^{(\mathrm{s})},
\end{align}
which leads to $Q_0^{(\mathrm{b})}\otimes Q_0^{(\mathrm{os})}=Q_0^{(\mathrm{bos})}$.

The second term in Eq. (\ref{eq:t_parameter}) depends on the bond angle through $\hat{R}_{\mu}\hat{R}_{\mu'}$.
By using the explicit expression for $\hat{\bm{R}}$ (Eq. (\ref{eq:R_trihelix})), we have 
\begin{align}
    \label{eq:spin-independent_os1}
    & \left[\cos^2\theta \left(\frac{\sqrt{3}}{3}Q_0^{(\mathrm{o})}-\frac{\sqrt{6}}{6}Q_u^{(\mathrm{o})}-\frac{\sqrt{2}}{2}Q_v^{(\mathrm{o})}\right) + \sin^2\theta \left(\frac{\sqrt{3}}{3}Q_0^{(\mathrm{o})}+\frac{\sqrt{6}}{3}Q_u^{(\mathrm{o})}\right) - \sin\theta\cos\theta \sqrt{2}Q_{yz}^{(\mathrm{o})} \right]\otimes Q_0^{(\mathrm{s})}\\
    \label{eq:spin-independent_os2}
    & \left[\frac{3}{4}\cos^2\theta \left(\frac{\sqrt{3}}{3}Q_0^{(\mathrm{o})}-\frac{\sqrt{6}}{6}Q_u^{(\mathrm{o})}+\frac{\sqrt{2}}{2}Q_v^{(\mathrm{o})}\right) + \frac{1}{4}\cos^2\theta \left(\frac{\sqrt{3}}{3}Q_0^{(\mathrm{o})}-\frac{\sqrt{6}}{6}Q_u^{(\mathrm{o})}-\frac{\sqrt{2}}{2}Q_v^{(\mathrm{o})}\right) \right.\nonumber \\
    & \left. + \sin^2\theta \left(\frac{\sqrt{3}}{3}Q_0^{(\mathrm{o})}+\frac{\sqrt{6}}{3}Q_u^{(\mathrm{o})}\right) \pm \frac{\sqrt{3}}{4}\cos^2\theta \sqrt{2}Q_{xy}^{(\mathrm{o})} + \frac{1}{2}\sin\theta\cos\theta \sqrt{2}Q_{yz}^{(\mathrm{o})} \pm \frac{\sqrt{3}}{2}\sin\theta\cos\theta \sqrt{2}Q_{zx}^{(\mathrm{o})} \right]\otimes Q_0^{(\mathrm{s})},
\end{align}
for the three bond, where the double signs are in the same order.
From these, the multipole corresponding to the second term in Eq. (\ref{eq:t_parameter}) is
\begin{align}
    \label{eq:spin-independent_bos}
    & \vcenter{\hbox{\includegraphics[width=1.2cm]{triangular_Q0_v3.pdf}}}\otimes \frac{1}{\sqrt{3}} \left( Q_0^{(\mathrm{o})} + \frac{1}{\sqrt{2}} (3\sin^2\theta-1)Q_u^{(\mathrm{o})} \right) \otimes Q_0^{(\mathrm{s})} \nonumber \\
    & + \cos^2\theta \left( \frac{1}{2}\vcenter{\hbox{\includegraphics[width=1.2cm]{triangular_Qx_v3.pdf}}} \otimes \frac{1}{\sqrt{2}}Q_v^{(\mathrm{o})} - \vcenter{\hbox{\includegraphics[width=1.2cm]{triangular_Qy_v3.pdf}}} \otimes \frac{\sqrt{6}}{4}Q_{xy}^{(\mathrm{o})} \right) \otimes Q_0^{(\mathrm{s})} \nonumber \\
    & + \sin 2\theta \left( \frac{1}{2}\vcenter{\hbox{\includegraphics[width=1.2cm]{triangular_Qx_v3.pdf}}} \otimes \frac{1}{\sqrt{2}}Q_{yz}^{(\mathrm{o})} - \vcenter{\hbox{\includegraphics[width=1.2cm]{triangular_Qy_v3.pdf}}} \otimes \frac{\sqrt{6}}{4}Q_{zx}^{(\mathrm{o})} \right) \otimes Q_0^{(\mathrm{s})}.
\end{align}
We notice that Eq. (\ref{eq:spin-independent_bos}) is symmetrically equivalent to Eqs. (\ref{eq:trihelix_multipoles2-1})-(\ref{eq:trihelix_multipoles2-3}).
Thus, the spin-independent hopping Hamiltonian can be decomposed into $Q_0^{(\mathrm{bos})}, Q_u^{(\mathrm{bos})}, Q_{3a}^{(\mathrm{bos})}$ and $G_u^{(\mathrm{bos})}$.

\section{List of results for $\bm{\lambda}_{l\mu;l'\mu'}^{\mathrm{(a)sym}}$ for the $d$ orbitals}
\label{app:d-list}
\subsection{$d$-$d$ orbitals}
For the symmetric potential case, there are four independent parameters:
\begin{align}
    K_{dd\sigma}:=
    \begin{pmatrix}
        &2&2& \\
        1&0&0&\overline{1}
    \end{pmatrix}_{\mathrm{sym}}, 
    K_{dd\pi}:=
    \begin{pmatrix}
        &2&2& \\
        1&1&0&\overline{2}
    \end{pmatrix}_{\mathrm{sym}}, \\
    K'_{dd\pi}:=
    \begin{pmatrix}
        &2&2& \\
        1&1&\overline{1}&\overline{1}
    \end{pmatrix}_{\mathrm{sym}}, 
    K'_{dd\delta}:=
    \begin{pmatrix}
        &2&2& \\
        1&2&\overline{1}&\overline{2}
    \end{pmatrix}_{\mathrm{sym}}.
\end{align}

\begin{align}
    \lambda_{xy;yz,y}^{\mathrm{sym}}&=-\sqrt{3}K_{dd\sigma}\hat{R}_y^2(1-\hat{R}_y^2)-K'_{dd\pi}\hat{R}_y^2(1-2\hat{R}_y^2)+\frac{K_{dd\pi}}{\sqrt{2}}(1-\hat{R}_y^2)(1-3\hat{R}_y^2)-K'_{dd\delta}\hat{R}_y^2(1-\hat{R}_y^2) \nonumber \\
    \lambda_{xy;yz,z}^{\mathrm{sym}}&=\hat{R}_y\hat{R}_z\{\sqrt{3}K_{dd\sigma}\hat{R}_y^2-K'_{dd\pi}(1-2\hat{R}_y^2)-\frac{K_{dd\pi}}{\sqrt{2}}(1-3\hat{R}_y^2)-K'_{dd\delta}(1-\hat{R}_y^2)\} \nonumber \\
    \lambda_{xy;yz,x}^{\mathrm{sym}}&=\hat{R}_x\hat{R}_y\{\sqrt{3}K_{dd\sigma}\hat{R}_y^2-K'_{dd\pi}(1-2\hat{R}_y^2)-\frac{K_{dd\pi}}{\sqrt{2}}(1-3\hat{R}_y^2)-K'_{dd\delta}(1-\hat{R}_y^2)\} \nonumber \\ \nonumber \\
    \lambda_{xy;3z^2-r^2,y}^{\mathrm{sym}}&=\hat{R}_y\hat{R}_z\{-\frac{K_{dd\sigma}}{2}(2+3(\hat{R}_x^2-\hat{R}_y^2))-\sqrt{3}K'_{dd\pi}(\hat{R}_x^2-\hat{R}_y^2)-\frac{\sqrt{6}}{4}K_{dd\pi}(2+3(\hat{R}_x^2-\hat{R}_y^2)) \nonumber \\
    &-\frac{\sqrt{3}}{2}K'_{dd\delta}(\hat{R}_x^2-\hat{R}_y^2)\} \nonumber \\
    \lambda_{xy;3z^2-r^2,z}^{\mathrm{sym}}&=(\hat{R}_x^2-\hat{R}_y^2)\{\frac{K_{dd\sigma}}{2}(1-3\hat{R}_z^2)-\sqrt{3}K'_{dd\pi}\hat{R}_z^2+\frac{\sqrt{6}}{4}K_{dd\pi}(1-3\hat{R}_z^2)-\frac{\sqrt{3}}{2}K'_{dd\delta}\hat{R}_z^2\} \nonumber \\
    \lambda_{xy;3z^2-r^2,x}^{\mathrm{sym}}&=\hat{R}_x\hat{R}_z\{\frac{K_{dd\sigma}}{2}(2-3(\hat{R}_x^2-\hat{R}_y^2))-\sqrt{3}K'_{dd\pi}(\hat{R}_x^2-\hat{R}_y^2)-\frac{\sqrt{6}}{4}K_{dd\pi}(-2+3(\hat{R}_x^2-\hat{R}_y^2)) \nonumber \\
    &-\frac{\sqrt{3}}{2}K'_{dd\delta}(\hat{R}_x^2-\hat{R}_y^2)\} \nonumber \\ \nonumber \\
    \lambda_{xy;x^2-y^2,y}^{\mathrm{sym}}&=\hat{R}_y\hat{R}_z\{\frac{\sqrt{3}}{2}K_{dd\sigma}(1-\hat{R}_z^2)+K'_{dd\pi}(1-\hat{R}_z^2)-\frac{\sqrt{2}}{4}K_{dd\pi}(1+3\hat{R}_z^2)-\frac{K'_{dd\delta}}{2}(1+\hat{R}_z^2)\} \nonumber \\
    \lambda_{xy;x^2-y^2,z}^{\mathrm{sym}}&=-\frac{\sqrt{3}}{2}K_{dd\sigma}(1-\hat{R}_z^2)^2+K'_{dd\pi}\hat{R}_z^2(1-\hat{R}_z^2)+\frac{\sqrt{2}}{4}K_{dd\pi}(1-\hat{R}_z^2)(1+3\hat{R}_z^2)-\frac{K'_{dd\delta}}{2}\hat{R}_z^2(1+\hat{R}_z^2) \nonumber \\
    \lambda_{xy;x^2-y^2,x}^{\mathrm{sym}}&=\hat{R}_x\hat{R}_z\{\frac{\sqrt{3}}{2}K_{dd\sigma}(1-\hat{R}_z^2)+K'_{dd\pi}(1-\hat{R}_z^2)-\frac{\sqrt{2}}{4}K_{dd\pi}(1+3\hat{R}_z^2)-\frac{K'_{dd\delta}}{2}(1+\hat{R}_z^2)\} \nonumber \\ \nonumber \\
    \lambda_{yz;3z^2-r^2,y}^{\mathrm{sym}}&=\hat{R}_x\hat{R}_y\{-\frac{K_{dd\sigma}}{2}(1+3\hat{R}_z^2)-\sqrt{3}K'_{dd\pi}\hat{R}_z^2+\frac{\sqrt{6}}{4}K_{dd\pi}(1-3\hat{R}_z^2)+\frac{\sqrt{3}}{2}K'_{dd\delta}(1-\hat{R}_z^2)\} \nonumber \\
    \lambda_{yz;3z^2-r^2,z}^{\mathrm{sym}}&=\hat{R}_x\hat{R}_z\{\frac{K_{dd\sigma}}{2}(1-3\hat{R}_z^2)-\sqrt{3}K'_{dd\pi}\hat{R}_z^2+\frac{3\sqrt{6}}{4}K_{dd\pi}(1-\hat{R}_z^2)+\frac{\sqrt{3}}{2}K'_{dd\delta}(1-\hat{R}_z^2)\} \nonumber \\
    \lambda_{yz;3z^2-r^2,x}^{\mathrm{sym}}&=\frac{K_{dd\sigma}}{2}(\hat{R}_y^2+2\hat{R}_z^2-3\hat{R}_x^2\hat{R}_z^2)-\sqrt{3}K'_{dd\pi}\hat{R}_x^2\hat{R}_z^2-\frac{\sqrt{6}}{4}K_{dd\pi}(\hat{R}_y^2+3\hat{R}_x^2\hat{R}_z^2)+\frac{\sqrt{3}}{2}K'_{dd\delta}\hat{R}_x^2(1-\hat{R}_z^2) \nonumber \\ \nonumber \\
    \lambda_{yz;x^2-y^2,y}^{\mathrm{sym}}&=\hat{R}_x\hat{R}_y\{\frac{\sqrt{3}}{2}K_{dd\sigma}(1-(2\hat{R}_y^2-\hat{R}_z^2))-K'_{dd\pi}(2\hat{R}_y^2-\hat{R}_z^2)+\frac{\sqrt{2}}{4}K_{dd\pi}(5-3(2\hat{R}_y^2-\hat{R}_z^2)) \nonumber \\
    &+\frac{K'_{dd\delta}}{2}(1-(2\hat{R}_y^2-\hat{R}_z^2))\} \nonumber \\
    \lambda_{yz;x^2-y^2,z}^{\mathrm{sym}}&=\hat{R}_x\hat{R}_z\{-\frac{\sqrt{3}}{2}K_{dd\sigma}(1+2\hat{R}_y^2-\hat{R}_z^2)-K'_{dd\pi}(2\hat{R}_y^2-\hat{R}_z^2)-\frac{\sqrt{2}}{4}K_{dd\pi}(1+3(2\hat{R}_y^2-\hat{R}_z^2)) \nonumber \\
    &+\frac{K'_{dd\delta}}{2}(1-(2\hat{R}_y^2-\hat{R}_z^2))\} \nonumber \\
    \lambda_{yz;x^2-y^2,x}^{\mathrm{sym}}&=\frac{\sqrt{3}}{2}K_{dd\sigma}(\hat{R}_y^2-\hat{R}_x^2(2\hat{R}_y^2-\hat{R}_z^2))-K'_{dd\pi}(2\hat{R}_y^2-\hat{R}_z^2)+\frac{\sqrt{2}}{4}K_{dd\pi}(\hat{R}_y^2-2\hat{R}_z^2-3\hat{R}_x^2(2\hat{R}_y^2-\hat{R}_z^2)) \nonumber \\
    &+\frac{K'_{dd\delta}}{2}\hat{R}_x^2(1-(2\hat{R}_y^2-\hat{R}_z^2)) \nonumber \\ \nonumber \\
    \lambda_{3z^2-r^2;x^2-y^2,y}^{\mathrm{sym}}&=\hat{R}_x\hat{R}_z\{K_{dd\sigma}(1-3\hat{R}_y^2)-2\sqrt{3}K'_{dd\pi}\hat{R}_y^2+\frac{\sqrt{6}}{2}K_{dd\pi}(1-3\hat{R}_y^2)-\sqrt{3}K'_{dd\delta}\hat{R}_y^2\} \nonumber \\
    \lambda_{3z^2-r^2;x^2-y^2,z}^{\mathrm{sym}}&=\hat{R}_x\hat{R}_y\{K_{dd\sigma}(1-3\hat{R}_z^2)-2\sqrt{3}K'_{dd\pi}\hat{R}_z^2+\frac{\sqrt{6}}{2}K_{dd\pi}(1-3\hat{R}_z^2)-\sqrt{3}K'_{dd\delta}\hat{R}_z^2\} \nonumber \\
    \lambda_{3z^2-r^2;x^2-y^2,x}^{\mathrm{sym}}&=\hat{R}_y\hat{R}_z\{K_{dd\sigma}(1-3\hat{R}_x^2)-2\sqrt{3}K'_{dd\pi}\hat{R}_x^2+\frac{\sqrt{6}}{2}K_{dd\pi}(1-3\hat{R}_x^2)-\sqrt{3}K'_{dd\delta}\hat{R}_x^2\} \nonumber
\end{align}

For the antisymmetric potential case, there are two parameters:
\begin{align}
    \tilde{K}_{dd\sigma}:=
    \begin{pmatrix}
        &2&2& \\
        1&0&0&\overline{1}
    \end{pmatrix}_{\mathrm{asym}},
    \tilde{K}_{dd\pi}:=
    \begin{pmatrix}
        &2&2& \\
        1&1&0&\overline{2}
    \end{pmatrix}_{\mathrm{asym}}.
\end{align}

\begin{align}
    \lambda_{xy;xy,y}^{\mathrm{asym}}&=\hat{R}_x\hat{R}_z\{2\sqrt{3}\tilde{K}_{dd\sigma}\hat{R}_y^2-\sqrt{2}\tilde{K}_{dd\pi}(1-\hat{R}_y^2)\} \nonumber \\
    \lambda_{xy;xy,z}^{\mathrm{asym}}&=\hat{R}_x\hat{R}_y(\hat{R}_x^2-\hat{R}_y^2)(2\sqrt{3}\tilde{K}_{dd\sigma}+\sqrt{2}\tilde{K}_{dd\pi}) \nonumber \\
    \lambda_{xy;xy,x}^{\mathrm{asym}}&=\hat{R}_y\hat{R}_z\{-2\sqrt{3}\tilde{K}_{dd\sigma}\hat{R}_x^2+\sqrt{2}\tilde{K}_{dd\pi}(1-\hat{R}_x^2)\} \nonumber \\ \nonumber \\
    \lambda_{xy;yz,y}^{\mathrm{asym}}&=(\hat{R}_x^2-\hat{R}_z^2)\{-\sqrt{3}\tilde{K}_{dd\sigma}\hat{R}_y^2+\frac{\tilde{K}_{dd\pi}}{\sqrt{2}}(1-\hat{R}_y^2)\} \nonumber \\
    \lambda_{xy;yz,z}^{\mathrm{asym}}&=\hat{R}_y\hat{R}_z\{\sqrt{3}\tilde{K}_{dd\sigma}(2\hat{R}_x^2-\hat{R}_y^2)+\frac{\tilde{K}_{dd\pi}}{\sqrt{2}}(1+2\hat{R}_x^2-\hat{R}_y^2)\} \nonumber \\
    \lambda_{xy;yz,x}^{\mathrm{asym}}&=\hat{R}_x\hat{R}_y\{-\sqrt{3}\tilde{K}_{dd\sigma}(2\hat{R}_z^2-\hat{R}_y^2)-\frac{\tilde{K}_{dd\pi}}{\sqrt{2}}(1+2\hat{R}_z^2-\hat{R}_y^2)\} \nonumber \\ \nonumber \\
    \lambda_{xy;3z^2-r^2,y}^{\mathrm{asym}}&=\hat{R}_y\hat{R}_z\{-\frac{\tilde{K}_{dd\sigma}}{2}(1+3(2\hat{R}_x^2-\hat{R}_z^2))+\frac{\sqrt{6}}{4}\tilde{K}_{dd\pi}(1-(2\hat{R}_x^2-\hat{R}_z^2))\} \nonumber \\
    \lambda_{xy;3z^2-r^2,z}^{\mathrm{asym}}&=(\hat{R}_x^2-\hat{R}_y^2)\{-\frac{\tilde{K}_{dd\sigma}}{2}(1-3\hat{R}_z^2)+\frac{\sqrt{6}}{4}\tilde{K}_{dd\pi}(1+\hat{R}_z^2)\} \nonumber \\
    \lambda_{xy;3z^2-r^2,x}^{\mathrm{asym}}&=\hat{R}_x\hat{R}_z\{\frac{\tilde{K}_{dd\sigma}}{2}(1+3(2\hat{R}_y^2-\hat{R}_z^2))-\frac{\sqrt{6}}{4}\tilde{K}_{dd\pi}(1-(2\hat{R}_y^2-\hat{R}_z^2))\} \nonumber \\ \nonumber \\
    \lambda_{xy;x^2-y^2,y}^{\mathrm{asym}}&=\hat{R}_y\hat{R}_z(3\hat{R}_x^2-\hat{R}_y^2)(\frac{\sqrt{3}}{2}\tilde{K}_{dd\sigma}+\frac{\sqrt{2}}{4}\tilde{K}_{dd\pi}) \nonumber \\
    \lambda_{xy;x^2-y^2,z}^{\mathrm{asym}}&=(\hat{R}_x^4-6\hat{R}_x^2\hat{R}_y^2+\hat{R}_y^4)(\frac{\sqrt{3}}{2}\tilde{K}_{dd\sigma}+\frac{\sqrt{2}}{4}\tilde{K}_{dd\pi}) \nonumber \\
    \lambda_{xy;x^2-y^2,x}^{\mathrm{asym}}&=\hat{R}_x\hat{R}_z(3\hat{R}_y^2-\hat{R}_x^2)(\frac{\sqrt{3}}{2}\tilde{K}_{dd\sigma}+\frac{\sqrt{2}}{4}\tilde{K}_{dd\pi}) \nonumber \\ \nonumber \\
    \lambda_{yz;3z^2-r^2,y}^{\mathrm{asym}}&=\hat{R}_x\hat{R}_y\{\frac{\tilde{K}_{dd\sigma}}{2}(1-9\hat{R}_z^2)+\frac{\sqrt{6}}{4}\tilde{K}_{dd\pi}(1-3\hat{R}_z^2)\} \nonumber \\
    \lambda_{yz;3z^2-r^2,z}^{\mathrm{asym}}&=-\hat{R}_x\hat{R}_z\{\frac{\tilde{K}_{dd\sigma}}{2}(1-3\hat{R}_z^2)+\frac{\sqrt{6}}{4}\tilde{K}_{dd\pi}(1-\hat{R}_z^2)\} \nonumber \\
    \lambda_{yz;3z^2-r^2,x}^{\mathrm{asym}}&=\frac{\tilde{K}_{dd\sigma}}{2}(-\hat{R}_y^2+\hat{R}_z^2+3\hat{R}_z^2(3\hat{R}_y^2-\hat{R}_z^2))-\frac{\sqrt{6}}{4}\tilde{K}_{dd\pi}(\hat{R}_y^2-\hat{R}_z^2-\hat{R}_z^2(3\hat{R}_y^2-\hat{R}_z^2)) \nonumber \\ \nonumber \\
    \lambda_{yz;x^2-y^2,y}^{\mathrm{asym}}&=\hat{R}_x\hat{R}_y\{\frac{\sqrt{3}}{2}\tilde{K}_{dd\sigma}(2-(3\hat{R}_x^2+\hat{R}_y^2))-\frac{\sqrt{2}}{4}\tilde{K}_{dd\pi}(3\hat{R}_x^2+\hat{R}_y^2)\} \nonumber \\
    \lambda_{yz;x^2-y^2,z}^{\mathrm{asym}}&=\hat{R}_x\hat{R}_z\{\frac{\sqrt{3}}{2}\tilde{K}_{dd\sigma}(\hat{R}_x^2-5\hat{R}_y^2)+\frac{\sqrt{2}}{4}\tilde{K}_{dd\pi}(2+\hat{R}_x^2-5\hat{R}_y^2)\} \nonumber \\
    \lambda_{yz;x^2-y^2,x}^{\mathrm{asym}}&=-\frac{\sqrt{3}}{2}\tilde{K}_{dd\sigma}(2\hat{R}_x^4+3\hat{R}_x^2\hat{R}_z^2-4\hat{R}_y^2\hat{R}_z^2-2\hat{R}_x^2+\hat{R}_y^2)-\frac{\sqrt{2}}{4}\tilde{K}_{dd\pi}(2\hat{R}_x^4+3\hat{R}_x^2\hat{R}_z^2-4\hat{R}_y^2\hat{R}_z^2-2\hat{R}_x^2-\hat{R}_y^2+2\hat{R}_z^2) \nonumber \\ \nonumber \\
    \lambda_{3z^2-r^2;3z^2-r^2,y}^{\mathrm{asym}}&=\hat{R}_x\hat{R}_z\{\sqrt{3}\tilde{K}_{dd\sigma}(1-3\hat{R}_z^2)+\frac{3}{\sqrt{2}}\tilde{K}_{dd\pi}(1-\hat{R}_z^2)\} \nonumber \\
    \lambda_{3z^2-r^2;3z^2-r^2,z}^{\mathrm{asym}}&=0 \nonumber \\
    \lambda_{3z^2-r^2;3z^2-r^2,x}^{\mathrm{asym}}&=\hat{R}_y\hat{R}_z\{-\sqrt{3}\tilde{K}_{dd\sigma}(1-3\hat{R}_z^2)-\frac{3}{\sqrt{2}}\tilde{K}_{dd\pi}(1-\hat{R}_z^2)\} \nonumber \\ \nonumber \\
    \lambda_{3z^2-r^2;x^2-y^2,y}^{\mathrm{asym}}&=\hat{R}_x\hat{R}_z\{\tilde{K}_{dd\sigma}(1-3\hat{R}_x^2)+\frac{\sqrt{6}}{2}\tilde{K}_{dd\pi}(1-\hat{R}_x^2)\} \nonumber \\
    \lambda_{3z^2-r^2;x^2-y^2,z}^{\mathrm{asym}}&=\hat{R}_x\hat{R}_y\{\tilde{K}_{dd\sigma}(1-3\hat{R}_z^2)-\frac{\sqrt{6}}{2}\tilde{K}_{dd\pi}(1+\hat{R}_z^2)\} \nonumber \\
    \lambda_{3z^2-r^2;x^2-y^2,x}^{\mathrm{asym}}&=\hat{R}_y\hat{R}_z\{\tilde{K}_{dd\sigma}(1-3\hat{R}_y^2)+\frac{\sqrt{6}}{2}\tilde{K}_{dd\pi}(1-\hat{R}_y^2)\} \nonumber \\ \nonumber \\
    \lambda_{x^2-y^2;x^2-y^2,y}^{\mathrm{asym}}&=\hat{R}_x\hat{R}_z\{\sqrt{3}\tilde{K}_{dd\sigma}(\hat{R}_x^2-\hat{R}_y^2)+\frac{\tilde{K}_{dd\pi}}{\sqrt{2}}(-2+\hat{R}_x^2-\hat{R}_y^2)\} \nonumber \\
    \lambda_{x^2-y^2;x^2-y^2,z}^{\mathrm{asym}}&=-\hat{R}_x\hat{R}_y(\hat{R}_x^2-\hat{R}_y^2)(2\sqrt{3}\tilde{K}_{dd\sigma}+\sqrt{2}\tilde{K}_{dd\pi}) \nonumber \\
    \lambda_{x^2-y^2;x^2-y^2,x}^{\mathrm{asym}}&=\hat{R}_y\hat{R}_z\{\sqrt{3}\tilde{K}_{dd\sigma}(\hat{R}_x^2-\hat{R}_y^2)+\frac{\tilde{K}_{dd\pi}}{\sqrt{2}}(2+\hat{R}_x^2-\hat{R}_y^2)\} \nonumber
\end{align}

\subsection{$s$-$d$ orbitals}
For the symmetric potential case, there is one parameter:
\begin{align}
    K_{sd\sigma}:=
    \begin{pmatrix}
        &0&2& \\
        1&0&0&\overline{1}
    \end{pmatrix}_{\mathrm{sym}}.
\end{align}

\begin{align}
    \lambda_{s;xy,y}^{\mathrm{sym}}&=K_{sd\sigma}\hat{R}_y\hat{R}_z \nonumber \\
    \lambda_{s;xy,z}^{\mathrm{sym}}&=K_{sd\sigma}(\hat{R}_x^2-\hat{R}_y^2) \nonumber \\
    \lambda_{s;xy,x}^{\mathrm{sym}}&=-K_{sd\sigma}\hat{R}_x\hat{R}_z \nonumber \\ \nonumber \\
    \lambda_{s;3z^2-r^2,y}^{\mathrm{sym}}&=-\sqrt{3}K_{sd\sigma}\hat{R}_x\hat{R}_z \nonumber \\
    \lambda_{s;3z^2-r^2,z}^{\mathrm{sym}}&=0 \nonumber \\
    \lambda_{s;3z^2-r^2,x}^{\mathrm{sym}}&=\sqrt{3}K_{sd\sigma}\hat{R}_y\hat{R}_z \nonumber \\ \nonumber \\
    \lambda_{s;x^2-y^2,y}^{\mathrm{sym}}&=K_{sd\sigma}\hat{R}_x\hat{R}_z \nonumber \\
    \lambda_{s;x^2-y^2,z}^{\mathrm{sym}}&=-2K_{sd\sigma}\hat{R}_x\hat{R}_y \nonumber \\
    \lambda_{s;x^2-y^2,x}^{\mathrm{sym}}&=K_{sd\sigma}\hat{R}_y\hat{R}_z \nonumber \\
\end{align}

For the antisymmetric potential case, there is one parameter:
\begin{align}
    \tilde{K}_{sd\sigma}:=
    \begin{pmatrix}
        &0&2& \\
        1&0&0&\overline{1}
    \end{pmatrix}_{\mathrm{asym}}.
\end{align}
The form of the hopping integral is the same as in the case of the symmetric potential and obtained by replacing $K_{sd\sigma}$ with $\tilde{K}_{sd\sigma}$.

\subsection{$p$-$d$ orbitals}
For the symmetric potential case, there are 4 parameters:
\begin{align}
    K_{pd\sigma}:=
    \begin{pmatrix}
        &1&2& \\
        1&0&0&\overline{1}
    \end{pmatrix}_{\mathrm{sym}}, 
    K_{pd\pi}:=
    \begin{pmatrix}
        &1&2& \\
        1&1&0&\overline{2}
    \end{pmatrix}_{\mathrm{sym}}, \\
    K_{pd\overline{\pi}}:=
    \begin{pmatrix}
        &1&2& \\
        1&\overline{1}&0&0
    \end{pmatrix}_{\mathrm{sym}}, 
    K'_{pd\pi}:=
    \begin{pmatrix}
        &1&2& \\
        1&1&\overline{1}&\overline{1}
    \end{pmatrix}_{\mathrm{sym}}.
\end{align}

\begin{align}
    \lambda_{y;xy,y}^{\mathrm{sym}}&=\hat{R}_z\{K_{pd\sigma}\hat{R}_y^2-\frac{K_{pd\pi}}{\sqrt{2}}(1-\hat{R}_y^2)+K'_{pd\pi}\hat{R}_y^2\} \nonumber \\
    \lambda_{y;xy,z}^{\mathrm{sym}}&=\hat{R}_y\{K_{pd\sigma}(\hat{R}_x^2-\hat{R}_y^2)+\frac{K_{pd\pi}}{\sqrt{2}}(1-\hat{R}_y^2)+\sqrt{3}K_{pd\overline{\pi}}\hat{R}_x^2+K'_{pd\pi}\hat{R}_z^2\} \nonumber \\
    \lambda_{y;xy,x}^{\mathrm{sym}}&=\hat{R}_x\hat{R}_y\hat{R}_z(-K_{pd\sigma}-\sqrt{3}K_{pd\overline{\pi}}+K'_{pd\pi}) \nonumber \\ \nonumber \\
    \lambda_{y;3z^2-r^2,y}^{\mathrm{sym}}&=-\sqrt{3}\hat{R}_x\hat{R}_y\hat{R}_z(K_{pd\sigma}+\frac{1}{\sqrt{2}}K_{pd\pi}+K'_{pd\pi}) \nonumber \\
    \lambda_{y;3z^2-r^2,z}^{\mathrm{sym}}&=\hat{R}_x\{\frac{\sqrt{6}}{4}K_{pd\pi}(1-\hat{R}_z^2)-\frac{K_{pd\overline{\pi}}}{2}(1-3\hat{R}_z^2)-\sqrt{3}K'_{pd\pi}\hat{R}_z^2\} \nonumber \\
    \lambda_{y;3z^2-r^2,x}^{\mathrm{sym}}&=\hat{R}_z\{\sqrt{3}K_{pd\sigma}\hat{R}_y^2-\frac{\sqrt{6}}{4}K_{pd\pi}(\hat{R}_x^2-\hat{R}_y^2)+\frac{K_{pd\overline{\pi}}}{2}(1-3\hat{R}_z^2)-\sqrt{3}K'_{pd\pi}\hat{R}_x^2\} \nonumber \\ \nonumber \\
    \lambda_{y;zx,y}^{\mathrm{sym}}&=-\hat{R}_y(\hat{R}_x^2-\hat{R}_z^2)(K_{pd\sigma}+\frac{K_{pd\pi}}{\sqrt{2}}+K'_{pd\pi}) \nonumber \\
    \lambda_{y;zx,z}^{\mathrm{sym}}&=\hat{R}_z\{-K_{pd\sigma}\hat{R}_y^2-\frac{K_{pd\pi}}{\sqrt{2}}(1-\hat{R}_z^2)+\sqrt{3}K_{pd\overline{\pi}}\hat{R}_x^2-K'_{pd\pi}(\hat{R}_x^2-\hat{R}_z^2)\} \nonumber \\
    \lambda_{y;zx,x}^{\mathrm{sym}}&=\hat{R}_x\{K_{pd\sigma}\hat{R}_y^2+\frac{K_{pd\pi}}{\sqrt{2}}(1-\hat{R}_x^2)-\sqrt{3}K_{pd\overline{\pi}}\hat{R}_z^2-K'_{pd\pi}(\hat{R}_x^2-\hat{R}_z^2)\} \nonumber \\ \nonumber \\
    \lambda_{y;x^2-y^2,y}^{\mathrm{sym}}&=\hat{R}_x\hat{R}_y\hat{R}_z(K_{pd\sigma}+\frac{K_{pd\pi}}{\sqrt{2}}+K'_{pd\pi}) \nonumber \\
    \lambda_{y;x^2-y^2,z}^{\mathrm{sym}}&=\hat{R}_x\{-2K_{pd\sigma}\hat{R}_y^2+\frac{\sqrt{2}}{4}K_{pd\pi}(\hat{R}_x^2-\hat{R}_y^2+2\hat{R}_z^2)+\frac{\sqrt{3}}{2}K_{pd\overline{\pi}}(\hat{R}_x^2-\hat{R}_y^2)+K'_{pd\pi}\hat{R}_z^2\} \nonumber \\
    \lambda_{y;x^2-y^2,x}^{\mathrm{sym}}&=\hat{R}_z\{K_{pd\sigma}\hat{R}_y^2-\frac{\sqrt{2}}{4}K_{pd\pi}(1+\hat{R}_z^2)-\frac{\sqrt{3}}{2}K_{pd\overline{\pi}}(\hat{R}_x^2-\hat{R}_y^2)+K'_{pd\pi}\hat{R}_x^2\} \nonumber \\ \nonumber \\
    \lambda_{z;3z^2-r^2,y}^{\mathrm{sym}}&=\hat{R}_x\{-\sqrt{3}K_{pd\sigma}\hat{R}_z^2+\frac{\sqrt{6}}{4}K_{pd\pi}(1-\hat{R}_z^2)+\frac{K_{pd\overline{\pi}}}{2}(1-3\hat{R}_z^2)\} \nonumber \\
    \lambda_{z;3z^2-r^2,z}^{\mathrm{sym}}&=0 \nonumber \\
    \lambda_{z;3z^2-r^2,x}^{\mathrm{sym}}&=\hat{R}_y\{\sqrt{3}K_{pd\sigma}\hat{R}_z^2-\frac{\sqrt{6}}{4}K_{pd\pi}(1-\hat{R}_z^2)-\frac{K_{pd\overline{\pi}}}{2}(1-3\hat{R}_z^2)\} \nonumber \\ \nonumber \\
    \lambda_{z;x^2-y^2,y}^{\mathrm{sym}}&=\hat{R}_x\{K_{pd\sigma}\hat{R}_z^2+\frac{\sqrt{2}}{4}K_{pd\pi}(\hat{R}_x^2-\hat{R}_y^2+2\hat{R}_z^2)-\frac{\sqrt{3}}{2}K_{pd\overline{\pi}}(\hat{R}_x^2-\hat{R}_y^2)-2K'_{pd\pi}\hat{R}_y^2\} \nonumber \\
    \lambda_{z;x^2-y^2,z}^{\mathrm{sym}}&=-2\hat{R}_x\hat{R}_y\hat{R}_z(K_{pd\sigma}+\frac{1}{\sqrt{2}}K_{pd\pi}+K'_{pd\pi}) \nonumber \\
    \lambda_{z;x^2-y^2,x}^{\mathrm{sym}}&=\hat{R}_y\{K_{pd\sigma}\hat{R}_z^2-\frac{\sqrt{2}}{4}K_{pd\pi}(\hat{R}_x^2-\hat{R}_y^2-2\hat{R}_z^2)+\frac{\sqrt{3}}{2}K_{pd\overline{\pi}}(\hat{R}_x^2-\hat{R}_y^2)-2K'_{pd\pi}\hat{R}_x^2\} \nonumber
\end{align}

For the antisymmetric potential case, there is one parameter:
\begin{align}
    \tilde{K}_{pd\sigma}:=
    \begin{pmatrix}
        &1&2& \\
        1&0&0&\overline{1}
    \end{pmatrix}_{\mathrm{asym}}, 
    \tilde{K}_{pd\pi}:=
    \begin{pmatrix}
        &1&2& \\
        1&1&0&\overline{2}
    \end{pmatrix}_{\mathrm{asym}}, \\
    \tilde{K}_{pd\overline{\pi}}:=
    \begin{pmatrix}
        &1&2& \\
        1&\overline{1}&0&0
    \end{pmatrix}_{\mathrm{asym}}, 
    \tilde{K}'_{pd\pi}:=
    \begin{pmatrix}
        &1&2& \\
        1&1&\overline{1}&\overline{1}
    \end{pmatrix}_{\mathrm{asym}}.
\end{align}
The form of the hopping integral is the same as in the case of the symmetric potential and obtained by replacing $K_{pd\sigma}, K_{pd\pi}, K_{pd\overline{\pi}}, K'_{pd\pi}$ with $\tilde{K}_{pd\sigma}, \tilde{K}_{pd\pi}, \tilde{K}_{pd\overline{\pi}}, \tilde{K}'_{pd\pi}$, respectively.

\end{widetext}

\bibliography{manuscript}

\end{document}